\documentclass[article]{aa}  

\usepackage{hyperref}
\hypersetup{unicode=true, colorlinks=true, linkcolor=magenta, citecolor=blue, filecolor=red, urlcolor=blue}

\usepackage{array}

\usepackage{graphicx}
\usepackage{subcaption}
\usepackage{txfonts}
\usepackage[utf8]{inputenc}
\usepackage{newunicodechar}
\newunicodechar{́}{\'}

\begin{document} 
   \title{Unveiling the faint diffuse X-ray emission in Westerlund\,1}
   \subtitle{EWOCS-VII}

\newcommand{\InstQMUL}{Astronomy Unit, School of Physics and Astronomy, Queen Mary University of London, London E1 4NS, UK}

\author{
  J.F. Albacete-Colombo\inst{1}
  \and M. Andersen\inst{2}
  \and M. De Becker\inst{3}
  \and J. Mackey\inst{4}
  \and C.J.K. Larkin\inst{5,6,7}
  \and M.G. Guarcello\inst{8}
  \and K. Anastasopoulou\inst{8,12}
  \and S. Sciortino\inst{8}
  \and E. Greco\inst{8}
  \and M. Miceli\inst{9,8}
  \and V. Sapienza\inst{8}
  \and E. Flaccomio\inst{8}
  \and A. Filócomo\inst{10,1}
  \and I.R. Stevens\inst{11}
  \and A. Bayo\inst{2}
  \and J.J. Drake\inst{12}
  \and M. Gennaro\inst{13,14}
  \and S.J. Gunderson\inst{15}
  \and F. Fraschetti\inst{12}
}

\institute{
  \inst{1} Universidad Nacional de Río Negro, Sede Atlántica, Don Bosco y Leloir, Viedma CP 8500, Río Negro, Argentina \\
  \inst{2} European Southern Observatory, Karl-Schwarzschild-Strasse 2, D-85748 Garching bei München, Germany \\
  \inst{3} Space Sciences STAR Institute, University of Liège, Quartier Agora, 19c, Allée du 6 Août, B5c, 4000 Sart Tilman, Belgium \\
  \inst{4} Dublin Institute for Advanced Studies, DIAS Dunsink Observatory, Dublin D15 XR2R, Ireland \\
  \inst{5} Max-Planck-Institut für Kernphysik, Saupfercheckweg 1, D-69117 Heidelberg, Germany \\
  \inst{6} Astronomisches Rechen-Institut, Heidelberg, Mönchhofstr. 12-14, D-69120 Heidelberg, Germany \\
  \inst{7} Max-Planck-Institut für Astronomie, Königstuhl 17, D-69117 Heidelberg, Germany \\
  \inst{8} INAF - Osservatorio Astronomico di Palermo G.S. Vaiana, Piazza del Parlamento 1,
  90134 Palermo, Italy \\
  \inst{9} Dipartimento di Fisica e Chimica E. Segrè, Università degli Studi di Palermo, Via Archirafi 36, 90123 Palermo, Italy\\
  \inst{10} Instituto de Astronomı́a y Fı́sica del Espacio (CONICET-UBA), CC. 67, Suc.28, C1428ZAA Buenos Aires, Argentina\\
  \inst{11} School of Physics and Astronomy, University of Birmingham, Edgbaston, Birmingham B15 2TT, UK \\
  \inst{12} Center for Astrophysics | Harvard \& Smithsonian, 60 Garden Street, Cambridge, MA 02138, USA \\
  \inst{13} Space Telescope Science Institute, 3700 San Martin Drive, Baltimore, MD 21218, USA \\
  \inst{14} The William H. Miller III Dept. of Physics \& Astronomy, Johns Hopkins University, Baltimore, MD 21218, USA \\
  \inst{15} Kavli Institute for Astrophysics and Space Research, MIT, 77 Massachusetts Ave., Cambridge, MA 02139, USA \\
}
   \date{}

\abstract
{
Westerlund 1 (Wd\,1) is the closest supermassive star cluster to the Sun, containing more than 100,000 stars of all spectral types and masses down to brown dwarfs. This population considerably heats the surrounding interstellar medium (ISM), making Wd\,1 a relevant place to investigate stellar feedback on intracluster gas.
}
{
We present the most detailed X-ray study to date of the diffuse emission in this region, aimed at providing the most favourable observational configuration for distinguishing and quantifying the point-source contribution to the true diffuse emission.
}
{
We analysed 36 \textit{Chandra} ACIS-I observations of Wd\,1 within an 8$\times$8 arcmin window. After removing 4922 point sources with ACIS Extract, using energy-dependent point spread function (PSF) models and adaptive 99\% enclosed-energy masks, we subtracted PSF-wing contamination from overlapping neighbours. It is critical in this crowded field to correct for background. We applied adaptive smoothing to reveal hot gas between stars in the soft\,(0.5-1.2 keV), medium\,(1.2-1.9 keV), and hard\,(1.9-7.0 keV) bands.
}
{
The spectral fitting suggests a shocked thermal plasma, with soft emission extending beyond the stellar core and diffuse X-rays correlating spatially with the massive stars. The cluster core (region~\#2) is best described by an APEC+PSHOCK model, capturing the continuous range of post-shock ionisation timescales produced by the many coexisting stellar wind shocks; it is characterised by collisionally excited plasma temperatures of $1.81\pm0.24$ and $4.11\pm1.98$~keV and the 6.7~keV FeK$_\alpha$ emission line. The outer region (Region~\#1) exhibits charge exchange emission (CXE) and is best described by a CXE+NEI model, with softer temperatures of $0.27\pm0.04$ and $1.09\pm0.05$~keV. Absorption column densities ($N_H$) range between 1.84 and 2.24 $\times10^{22}$~cm$^{-2}$ and the total (0.5--8.0 keV) diffuse luminosity of both regions is $\simeq 9.6\,(\pm0.3)\,\times\,10^{33}$~erg\,s$^{-1}$.
}
{
Near the core, the hard emission likely originates from thermalised Wolf-Rayet (WR) stellar winds and intense wind-wind collisions in regions of high massive-star density, while the adiabatic expansion of the hot plasma and turbulent mixing with the denser, cooler ISM ends up softening it. The soft diffuse component extends farther out and may be associated with CXE processes, spatially correlated with regions of lower extinction. No hint of a non-thermal contribution is revealed. The analytical cluster wind model, using the observed WR population, predicts $L_X^{\rm CWM} \simeq 2.1\,(\pm1.0)\times10^{34}$~erg~s$^{-1}$, a factor of $\sim$2.2 above the observed luminosity, consistent with the low-density, partially evacuated intracluster medium (ICM) inferred from the X-ray and infrared morphology. This low-density morphology at the cluster centre suggests that ultraviolet (UV) radiation can clear the ambient medium, producing an apparent outward displacement that likely reduces the $L_{\rm w}$-to-$L_{\rm X}$ conversion efficiency, measured as $\eta = L_{\rm X}/L_{\rm w} \simeq 4.3\times10^{-6}$, approximately
two orders of magnitude below that of the Cygnus~OB2 association and systematically
lower than every other massive star-forming region compiled here, resulting in
comparatively subluminous diffuse X-ray emission.
}

\titlerunning{Diffuse X-ray emission in Wd\,1}
\authorrunning{J. F. Albacete-Colombo et al.}
\maketitle

\section{Introduction}
Westerlund 1 (Wd\,1) is one of the most extreme starburst clusters in our galaxy, located 4.2$\pm$0.2 kpc away, making it the closest massive starburst to the Sun \citep{Negueruela2022}. Using the initial mass function (IMF) of \cite{Kroupa2002} and its known stellar content, approximately 1,500 M$_\odot$ of stars are contained within a radius of just 0.6 pc \citep{Clark2005}. Wd\,1 could have a total mass of 5$\times$10$^4$ to 10$^5$ M$_\odot$ \citep{Brandner2008}. This would make it one of the most massive and dense star clusters in the galaxy, far exceeding the stellar content of other known young clusters such as Quintuplet \citep{Quintuplet2012}, NGC 3603 \citep{ngc36032013}, and Arches \citep{Arches2018}.

Wd\,1 contains an extraordinary collection of evolved massive stars: 24 WR stars, over 80 blue supergiants, 6 yellow hypergiants, 9 red supergiants, and a luminous blue variable \citep{Negueruela2005}. The immense energy released into the interstellar medium (ISM) by massive stars through stellar winds generates dissipative shock waves in the local ISM \citep{Dorland1987}. Understanding the interaction between these winds and the local ISM is crucial for constraining stellar feedback models in general, superbubble models in particular, and for elucidating the mechanisms responsible for diffuse X-ray emission on spatial scales of several parsecs \citep{Rosen2014}.

Extensive observational campaigns have been undertaken to investigate the dissipation and redistribution of stellar wind energy in a variety of star-forming regions (SFRs), including the Rosette Nebula \citep{Townsley2003}, M17 \citep{Townsley2003}, and the Orion Nebula \citep{G2007}. However, conclusions regarding the astrophysical conditions of the diffuse X-ray emission can still be significantly affected by photon contamination from unresolved point sources. This contamination presents a major challenge in studying diffuse X-ray emission and complicates the interpretation of observational data. Moreover, diffuse X-ray emission provides a crucial observational diagnostic for constraining the average cluster wind velocity, which cannot significantly exceed the adiabatic sound speed within the cluster environment \citep{Rosen2014}.

Efforts to address this problem were made by \cite{Townsley2003}, who developed a sophisticated method to unveil diffuse X-ray emission from the combined X-ray emission of numerous resolved and unresolved young stars. In scenarios where the spatial distribution of stars extends over a few tens of arcminutes or even degrees across the sky, a reliable identification of the stellar population and an estimate of their contribution to the integrated X-ray luminosity of the cluster have been achieved. Most relevant studies have focused on the integrated X-ray emission from giant H II regions such as Carina (NGC 3372), M17, NGC 3576, NGC 3603, and 30 Doradus \citep{Townsley2003, Townsley2006, Townsley2009}. Two pioneering investigations of diffuse X-ray emission in young stellar environments have developed a new method for separating the contributions of unresolved stellar sources from those of the truly diffuse component. This challenge remains central to interpreting such observations and has established a more rigorous framework for addressing this persistent and complex separation problem. The first detailed analyses of a massive star cluster were conducted on Trumpler 16 \citep{Wolk2011}.

The first attempt to detect diffuse X-ray emission in the direction of Wd\,1 was made by \cite{Muno2006}, who used two \textit{Chandra} ACIS-S observations (OBSIDs 5411 and 6283) with total exposures of 18 and 42 ksec, respectively. They detected point-like X-ray sources up to a completeness limit of 2$\times$10$^{31}$ erg\,$\mathrm{s}^{-1}$, which represents the X-ray emission of massive stars in the region. This limit inadequately constrains the contribution of low-mass stars, leading to unresolved sources that largely contaminate the predicted diffuse emission. They identified diffuse emission of 3$\times$10$^{34}$ erg\,$\mathrm{s}^{-1}$ in the 2.0-8.0 keV range, but they did not analyse the 0.5-2.0 keV soft X-ray band, where most of the diffuse hot gas production is driven by the interaction of high-mass stellar winds with the surrounding ISM \citep{Canto2000}. More recently, \cite{Kavanagh2011} analysed the diffuse X-ray emission in the core of Wd\,1 using a 48 ksec \textit{XMM-Newton} observation. They found that the diffuse X-ray emission has a highly contaminated soft thermal component and a hard component that is either thermal or non-thermal. They used the X-ray spectra from the centre (2’ radius) with a He-like Fe 6.7 keV line emission. To avoid the influence of the two instrumental fluorescence lines at 1.49 and 1.75 keV, they restricted the spectral analysis to the 2-8 keV energy band. The assumption that all X-ray photons detected below 1.5 keV are probably foreground emissions is not necessarily correct. It is strongly influenced by the unresolved population of pre-main sequence stars and by the thermally interacting stellar winds of massive stars. These authors left the softer part of the diffuse emission unexplored and calculated an X-ray luminosity of 2$\times$10$^{33}$ erg\,s$^{-1}$ in the hard 2.0-8.0 keV range. The avoidance of the lower energies leads to a significant bias in the estimation of local $N_\mathrm{H}$ absorption in the diffuse X-ray emission. Consequently, the unabsorbed X-ray flux in the region is poorly constrained.

We overcome these limitations by analysing a deeper dataset consisting of 36 \textit{Chandra}/ACIS-I pointings that cover the full spatial extent of Wd\,1, with a total exposure time of 1.0 Ms. This represents a significant improvement in both sensitivity and angular resolution over previous studies and forms part of the Extended Westerlund One Chandra Survey (EWOCS; comprising two Chandra phases and complementary JWST observations)\footnote{https://westerlund1survey.wordpress.com/}, described in detail by \cite{Guarcello2019cxo}.

In this study, we analyse the morphology and spectral properties of the large-scale diffuse X-ray emission in Wd\,1 and investigate its possible origins. We derive key astrophysical parameters of the diffuse structures across both large and small spatial scales within the region.

\section{The EWOCS data}
\label{observations}

\begin{figure}[ht!]
\centering
\includegraphics[width=9cm,angle=0]{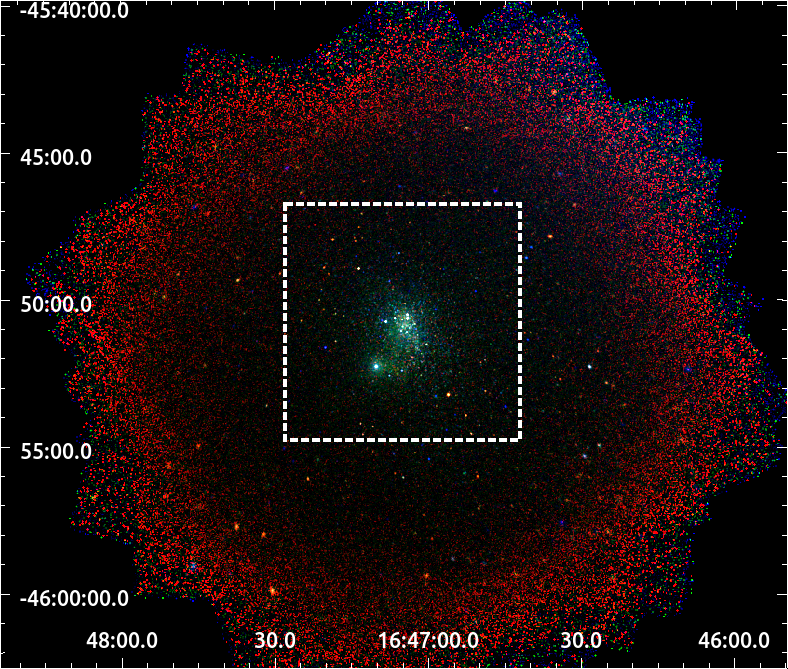}
\caption{\small Stacked projected observation of X-ray flux in units of ph\,cm$^{-2}$\,s$^{-1}$\,arcsec$^{-2}$. Colour-coded energy bands are soft [0.5-1.2] keV (red), medium [1.2-1.9] keV (green), and hard [1.9-7.0] keV (blue). Dashed white lines refer to an 8'$\times$8' arcmin size centred at RA: 16:47:04.8 and DEC: -45:50:48.1.}
\label{fig:img_fov}
\end{figure}

The EWOCS project obtained 36 \textit{Chandra} ACIS-I pointings with a total exposure time of 1.0 Ms in the Wd 1 region. Each observation with its aim point was adjusted according to the nominal roll angle to ensure complete coverage of the cluster core without gaps. This provided sub-arcsecond spatial resolution and high sensitivity in the central part of the ACIS-I detector, enabling detailed observation of the cluster core. All observations were acquired in VFAINT (5$\times$5 pixel island) mode, which is optimal for identifying and removing events originating from the endpoints of particle event tracks that cannot be eliminated using standard grade analysis based on 3$\times$3 (FAINT) islands. This set-up improves the separation of genuine X-ray diffuse emission photons from point-like detected X-ray sources, instrumental noise, and photons from the stellar background. It allows the exploration of low-surface-brightness extended features and the identification of diffuse emission patterns. 

The EWOCS survey was performed on a single RA=16:46:58.68  and DEC=-45:51:22.0 nominal pointing, and the description of the survey design, observational strategy, analysis of the observations, identification of sources, and source catalogue are discussed in detail by \cite{Guarcello2023}. 
All the observations were uniformly reprocessed using version 4.15 of the CIAO software \citep{Fruscione2014}, combined with CALDB 4.10.4 calibration database files (see Figure \,\ref{fig:img_fov}). To get consistency in the calibration procedure, we re-ran the Level 1 to Level 2 processing of event files using the CIAO \texttt{chandra\_repro} meta-task. Notably, we checked that the \textsc{vfaint\_pha=yes} option was set to flag bad events and filter them by events of \textsc{status==b0}. All the observations were processed with newer gain files, and all calibration files were updated during the \textsc{acis\_process\_event} to the latest available gain file version compatible with the observation.

All exposure maps for the observations, including the calibration background files, were created with the task \textsc{ae\_make\_emaps}. The mean energy that best represents the observed X-ray emission in the 0.5–7.0 keV band is 1.9 keV, so all exposure-corrected broadband images have the most accurate flux corrections. For the analysis of the three energy ranges: soft [0.5--1.2], medium [1.2--1.9], and hard [1.9--7.0] keV. We calculated exposure maps at 0.9, 1.5, and 2.3 keV, respectively. In addition, all exposure maps have a resolution of 1$\arcsec/pixel$, as required for diffuse emission studies by the ACIS Extract (AE; \citealt{Broos2012}) software. 
Vignetting in the images is corrected by dividing the raw count maps by these energy-dependent exposure maps, which encode the spatially varying effective area of the telescope–detector combination across the field of view. The particle-induced background in the images was estimated and subtracted using the ACIS stowed-position background calibration files acquired with the detector shielded from celestial X-rays, normalised to match each observation in the 7–12 keV band where astrophysical emission is negligible \citep{Hickox2006}. This procedure is described in detail in Appendix~\ref{ap:analysis}.

In appendix \ref{ap:xray_data} we describe the analysis performed to our X-ray observations and discuss the impact of active galactic nucleus (AGN) background contamination (Section \ref{sec:agn}), foreground contribution (Section \ref{sec:foreground}), and the impact of undetected sources on our diffuse emission analysis (Section \ref{sec:bkg}).

\section{Diffuse emission and imaging analysis}\label{sec:smooth}

We began the analysis of our 1\,Ms \texttt{Chandra} ACIS-I X-ray observation by combining all available observations to improve photon statistics. However, merged event files are unsuitable for accurate background subtraction or spectral analysis, as they do not preserve observation-specific aspect solutions required for reliable point spread function (PSF) modelling and response file generation. Consequently, image and spectral analyses must be performed separately for each observation to ensure accurate extraction regions and calibration products, Pulse Height Amplitude (PHA), Response Matrix Files (RMF), and Ancillary Response Files (ARF). Only after this step the results can be combined to generate smoothed maps with a defined S/N (see Sections \ref{sec:spectral} and \ref{sec:single}). Our methodology for dealing with diffuse emission is described in Appendix\,\ref{ap:analysis}.\\

A key challenge in analysing diffuse emission is determining the statistical significance of the observed structures, which strongly depends on the intrinsic diffuse emission and the smoothing method used. To investigate possible diffuse features within and outside the region, we used the task \textsc{tara\_smooth} in the AE software \citep{Broos2012}, which employs adaptive kernel smoothing at specified S/N ratios. We assumed values of 9, 12, 14, and 16, resulting in different smoothing radii. The choice of S/N ratio critically affects the ability to reveal or suppress true X-ray structures across spatial scales. The analysis performed here is the same as we performed and described in detail in \cite{Albacete-Colombo2023a}.

\begin{figure*}[ht!]
\centering
\includegraphics[width=9.1cm,angle=0]{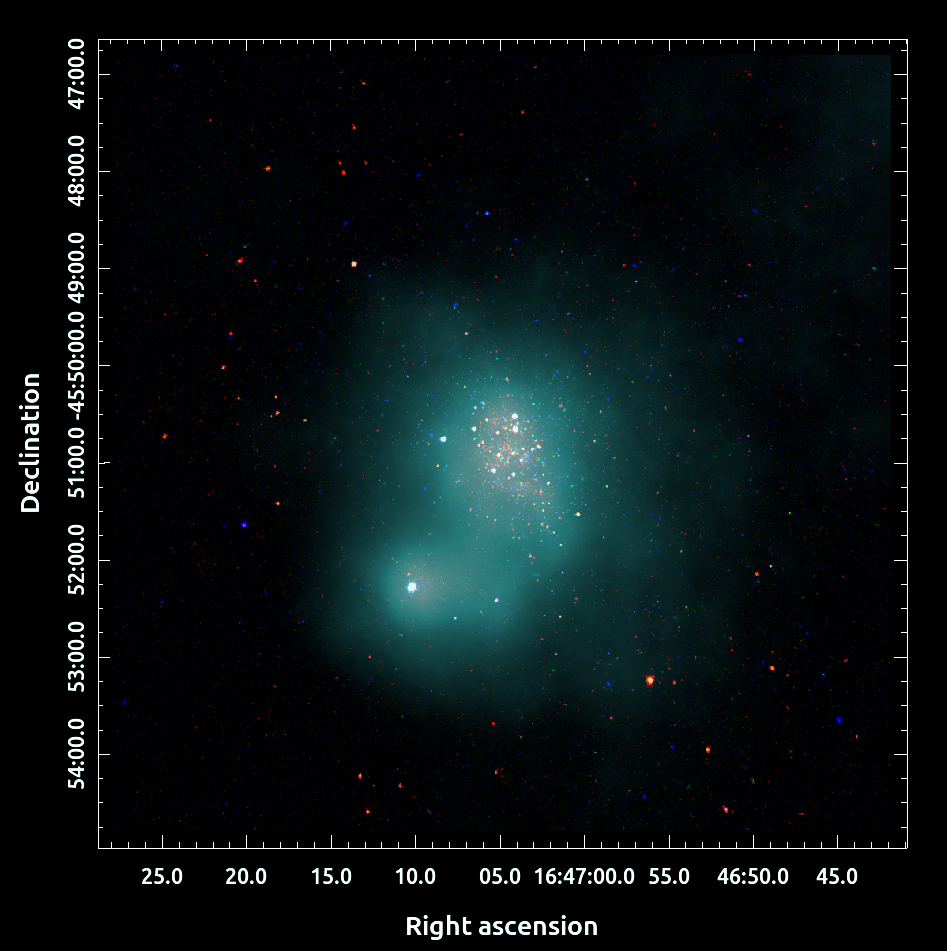}
\includegraphics[width=9.1cm,angle=0]{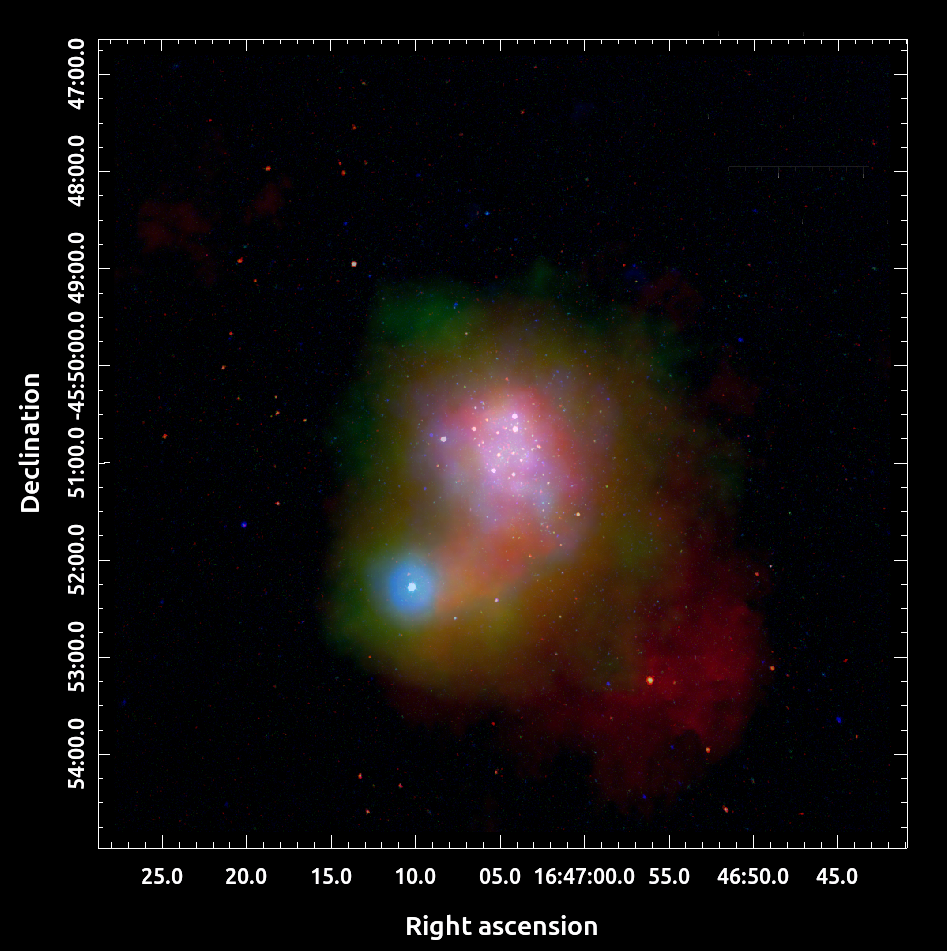}
\caption{\small Left: Diffuse X-ray emission in Alice Blue colour in the 0.5-7.0 keV energy band and superimposed on the stacked observation with point X-ray sources. Right: Colour-coded X-ray images in the soft ($0.5-1.2~\mathrm{keV}$; red), medium ($1.2-1.9~\mathrm{keV}$; green), and hard ($1.9-7.0~\mathrm{keV}$; blue) bands. The X-ray emission of the point sources is also shown for the same energy bands.}
\label{img_ph_fov}
\end{figure*}

Our analysis focused on an $8 \times 8$~arcmin$^2$ region that encompasses the dense stellar population of Wd\,1. To examine the spatial distribution of the X-ray emission, we generated a smoothed flux map with a resolution of $1024 \times 1024$ pixels. Two key parameters governed the smoothing procedure: (i) the S/N, which defines the statistical significance required for the smoothed flux, and (ii) the smoothing radius, which was adjusted to ensure a uniform spatial resolution across different energy bands. The maximum allowed smoothing radius was constrained to 142 pixels, within which the S/N condition was restricted.

Smoothing was initially performed on the full $0.5$--$7.0$~keV band. After testing multiple S/N thresholds (9, 12, 14, and 16), we adopted S/N = 14 as an optimal compromise between spatial resolution and the detection of genuine diffuse structures on arcminute scales. The resulting flux maps are expressed in units of photons\,cm$^{-2}$\,s$^{-1}$\,arcsec$^{-2}$. For the  sub-bands $-$ soft ($0.5$--$1.2$~keV), medium ($1.2$--$1.9$~keV), and hard ($1.9$--$7.0$~keV) $-$ we employed adaptive smoothing using radius kernels derived from the band with the lowest photon statistics (soft band), maintaining an approximate S/N of 14 to avoid over-smoothing, especially near detector edges. For medium and hard bands with higher photon counts, the smoothing radius map of the soft band was imposed via the \textsc{fixed\_radius\_map\_fn} parameter within the \textsc{tara\_smooth} routine, ensuring consistent spatial scales across bands while achieving higher S/N levels.
Figure~\ref{img_ph_fov} (left) shows the morphology of the diffuse emission in the full $0.5$--$7.0$~keV band, while Figure~\ref{img_ph_fov} (right) presents the smoothed maps in the soft, medium, and hard sub-bands within the same FOV. These energy bands were chosen to separate distinct physical emission components; notably, the soft band is sensitive to absorption and emission effects (see Section~\ref{sec:spectral}).

The bright X-ray magnetar CXO~164710.2--455217, highlighted in blue in Fig.~\ref{img_ph_fov}-right, contributes significantly to the apparent extended emission beyond $\sim$1.5 times its nominal PSF radius. This feature results from the scattering of X-ray photons by surrounding gas and dust structures \citep{Corral2011}, and it must be carefully distinguished from the intrinsic diffuse emission of Wd\,1. A precise characterisation of the spatial extent of the diffuse X-ray emission is therefore essential to disentangle these components, allowing us to assess the relative importance of hot plasma generated by stellar wind shocks in the cluster core versus scattered light from the magnetar in shaping the observed diffuse emission.

\section{Spatial modelling}
The spatial distribution of the hot X-ray-emitting gas within the interstellar medium (ISM) reveals the gas dynamics that produce diffuse X-ray emission and how this hot plasma coexists with the colder ambient gas of young star-forming regions (SFRs). We characterise it here through its radial surface-brightness profile (Sect.~\ref{sec:radprofile}) and a hardness-ratio map (Sect.~\ref{sec_hr}).

\subsection{Radial surface-brightness profile}
\label{sec:radprofile}

The stellar population of Wd\,1, both in the optical and in X-rays, exhibits a well-defined radial distribution with a clear decline at distances of $\sim$2-3 pc from the cluster core \citep{Clark2005}. The diffuse X-ray emission extends into the inner regions. We observed some spatial structures that are difficult to reproduce with simple analytical models \citep{Townsley2003, Albacete-Colombo2023a}. In addition, the large heliocentric distance of Wd\,1 relative to other massive star-forming regions (e.g., Carina, Cygnus OB2) limits the spatial resolution of small-scale features in its diffuse X-ray emission.

To quantify the spatial extent of the hot plasma, we computed the radial surface-brightness profile of the diffuse X-ray emission using concentric annuli centred at RA=16:47:04.532, DEC=$-$45:50:49.84. Assuming a cluster distance of $d = 4,250$~pc, the profile was expressed in both angular ($r$ [arcmin]) and physical ($R$ [pc]) units. 

\begin{figure*}[ht!]
\centering
\includegraphics[width=8.6cm]{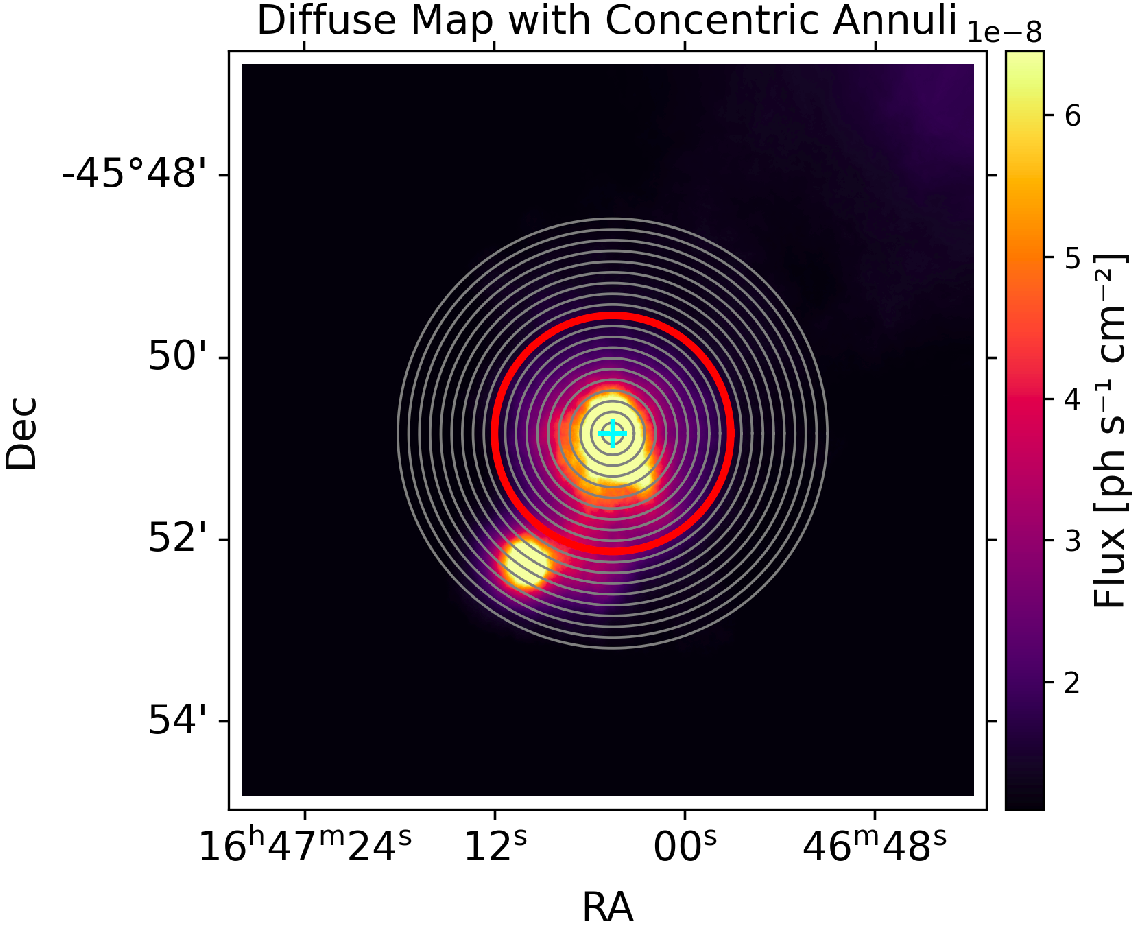}
\includegraphics[width=9.7cm]{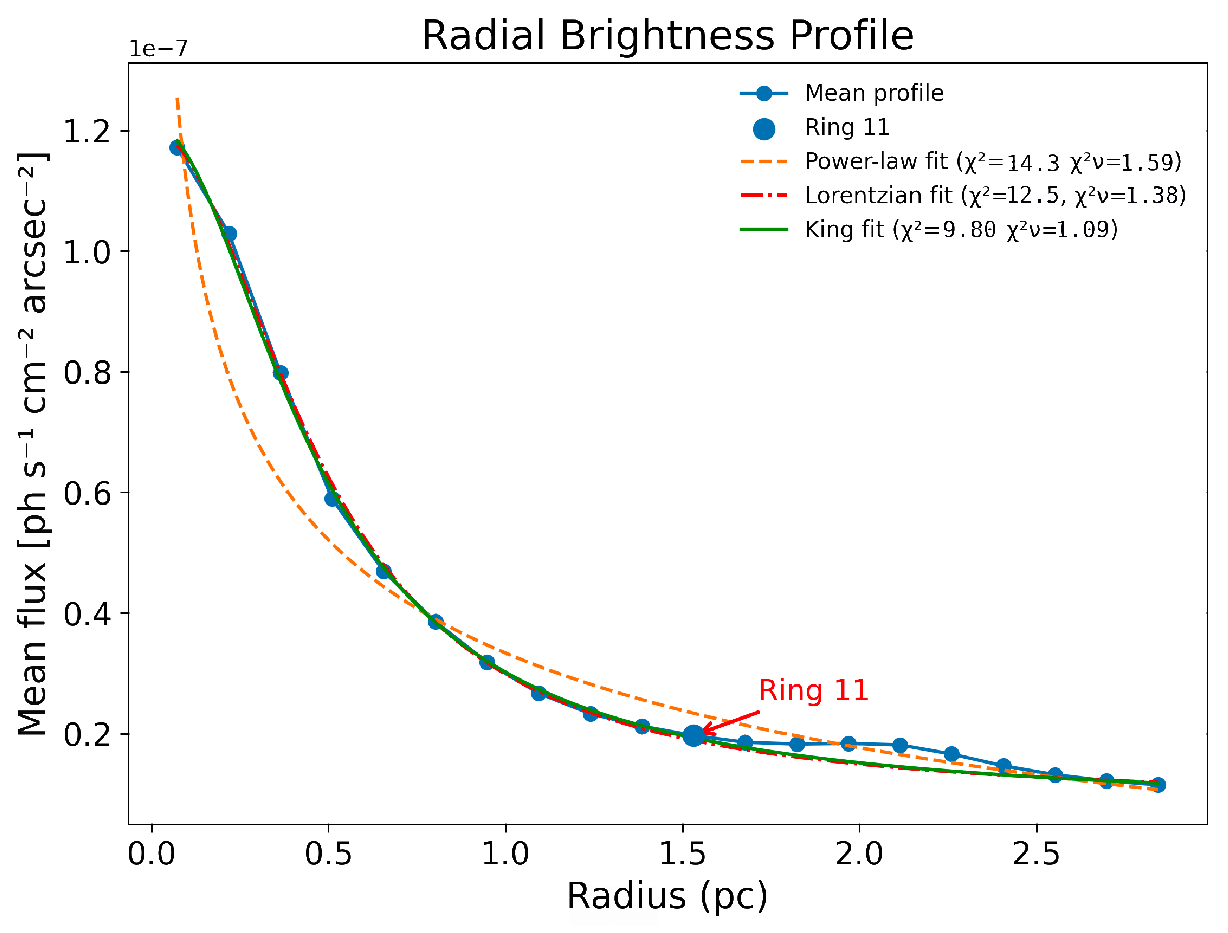}
\caption{\small 
Left: \textit{Chandra} image of Wd\,1 in the 0.5–7.0 keV band with concentric annuli overplotted, used to extract the radial surface brightness profile. The red circle with a radius of 1.5 pc marks the extent of the diffuse emission until the background level is reached. 
Right: Extracted radial profile of the diffuse X-ray emission with best-fit analytical models: power-law (blue), Lorentzian (red), and King profile (green). The King profile provides the best description of the data, with a reduction of $\chi^2_{\nu}$ close to unity. A small excess between 1.5 and 2.5 parsecs is associated with the scattering halo of X-ray photons from the magnetar CXO~164710.2$-$455217.}
\label{fig:img_rad}
\end{figure*}

The extracted distribution was fitted with three functional forms commonly used to describe diffuse X-ray emission and scattering halos: a power-law, a King profile, and a Lorentzian function (Figure~\ref{fig:img_rad}). The resulting best-fit parameters are listed below.

\begin{itemize}
    \item Power-law: $I(r) = A\,r^{-\alpha} + C$, with $A = (9.68 \pm 4.67) \times 10^{-8}$, $\alpha = 0.26 \pm 0.11$, and $C = (-6.35 \pm 4.45) \times 10^{-8}$. The fit yields $\chi^2 = 14.3$ ($\chi^2_\nu = 1.59$). The shallow slope indicates only a mild radial decline, inconsistent with the steep decrease expected for scattering-dominated halos, and therefore does not provide a physically meaningful description of the diffuse emission.
     
    \item Lorentzian: $I(r) = I_{0}\,\dfrac{\gamma^{2}}{(r-r_{0})^{2} + \gamma^{2}} + C$, with $I_{0} = (1.11 \pm 0.19) \times 10^{-7}$, $r_{0} \approx 0.0 \pm 0.17$~pc, $\gamma = 0.48 \pm 0.10$~pc, and $C = (8.80 \pm 2.30) \times 10^{-9}$. The fit yields $\chi^2 = 12.5$ ($\chi^2_\nu = 1.38$). Although this function reproduces the profile with a characteristic width of $\sim$0.5~pc, it is better interpreted as reflecting broadening processes that produce extended wings in the distribution, such as small-angle scattering by interstellar dust or turbulent shock interactions between stellar winds and the ISM.

    \item King profile: $I(r) = I_{0}\,[1+(r/r_{c})^{2}]^{-\beta} + C$, with $I_{0} = (1.13 \pm 0.14) \times 10^{-7}$, $r_{c} = 0.43 \pm 0.23$~pc, $\beta = 0.87 \pm 0.52$, and $C = (7.65 \pm 5.78) \times 10^{-9}$. The fit yields $\chi^2 = 9.8$ ($\chi^2_\nu = 1.09$). The compact core radius and moderate outer slope provide the best overall representation, reproducing both the central concentration and the gradual decline of the diffuse halo. Physically, this suggests that the hot gas is at least partially confined by the cluster’s gravitational potential, consistent with a compact stellar core and a density distribution approaching hydrostatic equilibrium.
\end{itemize}

Among the tested models, the King profile provides the most reliable representation of the diffuse emission in Wd\,1. It is characterised by a reduced chi-square close to unity and a physically motivated structure. 

The diffuse X-ray emission profile derived here can be directly compared with the stellar mass distribution in Wd\,1 to assess whether the cluster potential gravitationally confines the hot gas. Recent HST proper motion studies by \citet{Wei2025} fit an Elson-Fall-Freeman (EFF) profile to the stellar density, obtaining a core radius $r_c \approx 0.40$-$0.50$~pc for the full cluster sample, with mass segregation producing smaller core radii ($\sim$0.3~pc) for higher-mass stars. This is consistent with earlier estimates by \citet{Brandner2008}, who reported a stellar core radius of $\sim$0.4~pc (corresponding to $\sim$36\arcsec\ at $d \approx 3.7$--$4.0$~kpc). Our best-fit King profile for the diffuse X-ray emission yields $r_c = 0.43 \pm 0.23$~pc and $\beta=0.87\pm0.52$ reveals a compact core ($R_c\lesssim0.5$~pc) surrounded by an extended halo. The excellent agreement between the X-ray and stellar core radii indicates that the hot plasma is partially confined by the gravitational potential of the stellar core, consistent with hydrostatic equilibrium near the cluster centre. However, the shallow slope parameter $\beta$, is significantly lower than the outer power-law slope of the stellar EFF profile ($\gamma\sim2$--$3$), which indicates that the outer regions ($0.5\lesssim R \lesssim1.5$~pc) are shaped by additional physical processes beyond pure gravitational confinement. So, the combined action of wind-ISM interactions acts to disperse the hot gas and create the gradually declining diffuse halo observed at larger radii. This multi-scale picture supports a scenario in which thermal processes dominate the central gravitational well. At the same time, the extended X-ray envelope reflects the interplay between stellar feedback and the surrounding ISM.

Since most of the X-ray emission originates from the core region of Wd\,1, rigorous separation from contaminating sources is essential. This issue is particularly acute for the magnetar CXO~164710.2$-$455217, whose scattered photons have previously been included in measurements of the diffuse emission. This contamination artificially inflates the inferred spatial extent of the diffuse component to approximately 2.2~arcmin \citep{Muno2006, Kavanagh2011}, resulting in a fundamental mischaracterisation of the true diffuse emission. Consequently, the apparent contribution of the magnetar arises solely from photon scattering and is not related to cluster-driven processes. In the following section, we examine this scattering mechanism in detail to robustly isolate and quantify its contribution.

\subsection{Hardness ratio map}
\label{sec_hr}

One of the most common approaches to characterising spatial variations in X-ray spectra is through a hardness ratio (HR) analysis. This method is particularly useful in studies of extended X-ray sources, especially those involving diffuse emission. We generated HR maps by combining images from different energy bands, which allows us to estimate the characteristic photon energy while preserving spatial resolution. Specifically, we produced diffuse emission maps in the soft (S: 0.5–1.9~keV) and hard (H: 1.9–7.0~keV) bands and defined the hardness ratio as $\mathrm{HR} = (H-S)/(H+S)$, following the standard convention in which harder sources yield positive HR values. The HR map was constructed from smoothed images following the procedure described in Section~\ref{sec:smooth}. The smoothing kernel radii were derived from the corresponding soft-band image with  S/N~$\geq$~14, and the same smoothing scale was subsequently applied to the hard-band image. This approach minimises artificial diffuse structures that might otherwise arise from using mismatched smoothing radii \citep{Albacete-Colombo2023a}.

\begin{figure}[ht!]
\includegraphics[width=9.3cm, height=7.9cm, angle=0]{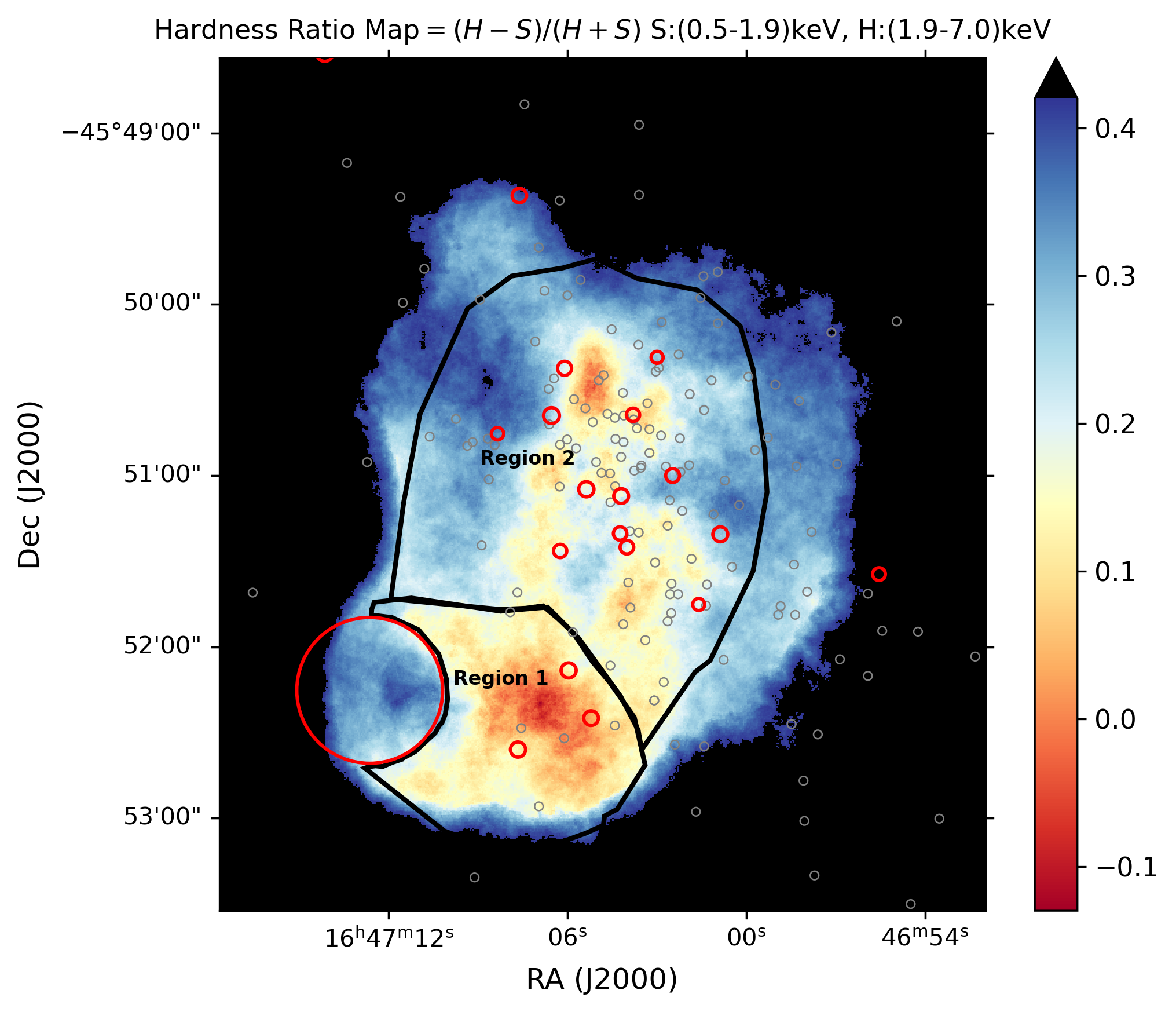}
\caption{\small
HR diffuse emission map derived from the soft (0.5--1.9 keV) and hard (1.9--7.0 keV) bands using the convention $\mathrm{HR} = (H-S)/(H+S)$, so that harder regions have positive values and softer regions have negative values. The colour scale ranges from red (soft, $\mathrm{HR} \approx -0.13$) through yellow to dark blue (hard, $\mathrm{HR} \geq 0.5$), with the palette saturating at $\mathrm{HR} = 0.42$. Pixels where the total (0.5--7.0 keV) flux falls below $1.4\times10^{-8}$ ph cm$^{-2}$ s$^{-1}$ arcsec$^{-2}$ are blended steeply towards the blue end of the scale, reflecting decreasing signal-to-noise rather than intrinsic spectral hardness. Small grey circles mark the locations of massive stars, while the red circles indicate evolved massive stars (supergiants and Wolf-Rayet stars). The two black polygons delineate Regions~\#1 and \#2 selected for spectral extraction (see Section~\ref{sec:spectral}). The large red circle at the south-eastern edge of region~\#1 has a radius of 26.5~arcsec and is centered on the magnetar CXO~164710.2$-$455217; this area is excluded from the diffuse emission analysis, and the choice of its radius is justified in Appendix~\ref{sec:cxo} (Fig.~\ref{fig:mag_radprof}).}
\label{img_hr}
\end{figure}

A key challenge when interpreting HR maps is to define the threshold above which individual X-ray photons from background sources begin to dominate over the truly diffuse emission. Based on simulated power-law spectra with photon $\Gamma$-index between 1.0 and 1.3 \citep{Alexander2003}, the expected HR values for AGNs are higher than $0.4$. In our map, such values indicate significant contamination from obscured AGNs emitting below the observational detection limit, which collectively contribute to the hard X-ray background (i.e., above 3~keV) \citep{May2008}. A quantitative assessment of the AGN contamination fraction, including an estimate of its contribution to the diffuse X-ray flux, is provided in Appendix~\ref{sec:agn}. Consequently, the HR map reveals genuine spatial variations in the hardness of the diffuse emission, reflecting a complex mixture of astrophysical components and emission mechanisms (see Figure~\ref{img_hr}).

We detected diffuse soft emission in regions beyond the cluster core extension (approximately 1.5 arcmin), where the density of massive stars becomes low. This area was not considered in previous works \citep{Muno2006, Kavanagh2011} due to the limited sensitivity of available observations. We defined two extraction regions optimised to maximise the S/N  of the diffuse emission spectra, enabling spatially resolved spectral analysis (see section~\ref{sec:spectral}). In region~\#1, which contains only six massive stars (3 MS + 3 evolved), the emission is primarily soft and extends a few tenths of a parsec from these stars. We constrained region~\#1 to avoid contamination from scattered photons originating from the magnetar CXO~164710.2-455217 (see section~\ref{sec:cxo}), and we accounted for the contribution of undetected faint cluster members as quantified in Appendix~\ref{sec:bkg}. In contrast, the diffuse emission in region~\#2 is dominated by medium-energy photons near the concentration of massive stars but becomes weaker beyond the cluster core.

We quantified the spatial distribution of HR values by defining two distinct zones:  
\begin{itemize}
\item Region 1: $-0.17 \leq \mathrm{HR} < 0.11$, with a median value of $-0.09 \pm 0.05$. This area is predominantly characterised by soft diffuse emission and covers approximately $1.47$ arcmin$^2$. Moving outward from the diffuse emission, the median HR increases to about $0.30$, indicating spectral hardening likely caused by the increasing relative contribution of the hard (stellar/instrumental) X-ray background, which becomes dominant in the measured signal.
\item Region 2: $-0.01 \leq \mathrm{HR} < 0.27$, with a median value of $0.21 \pm 0.08$. This region extends to $4.45$ arcmin$^2$, approximately three times larger than region \#1. Its harder nature may be related to intermediate-to-high spectral hardening, potentially linked to the central concentration of massive stars, which produce shocked ISM at higher temperatures, or to the presence of additional non-thermal emission at greater distances from the massive stellar population.  
\end{itemize}

The areas, HR ranges, and their median values were determined by counting the number of pixels and their HR values inside each of the diffuse emission masks. The quoted uncertainties on the median HR ($\pm0.05$ for region~\#1 and $\pm0.08$ for region~\#2) represent the standard error of the median, computed from the pixel-to-pixel dispersion of HR values within each mask divided by the square root of the number of independent resolution elements. These results suggest differences in the underlying astrophysical processes shaping the observed emission, which is further examined in Section~\ref{sec:spectral}.

\section{Spectral analysis of diffuse emission}
\label{sec:spectral}

X-ray spectral fitting of diffuse emission requires a sufficient number of X-ray photons for meaningful constraints. Increasing the source extraction areas is the most direct way to gain higher S/N. According to our analysis of the HR map and colour energy view of the region, we have identified two separate sources of emissions. These probably originate from different astrophysical conditions, so we decided to define the extraction regions based on these differences. This is more physically motivated than a simple S/N condition.

\subsection{Spectral extraction}
We focus on two distinct regions of Figure \ref{img_hr} and paid particular attention to the recognisable signs of hardness changes between them. This helps characterise the different astrophysical processes that occur on large spatial scales. The first region (region \#1), located south of Wd 1, shows weaker emission dominated mainly by X-rays in the 0.5 to 1.9 keV range. The total number of X-ray photons in the spectrum is 4200. We suspect low-temperature processes, such as thermal emission from weak stellar-wind shocks in the dense ISM or charge-exchange emission (CXE) at the interface between hot and cool gas phases. Extraction region \#1 covers 1.47 arcminutes$^2$ on the sky, which results in large response matrices for the spectral analysis. This region avoids the influence of scattered hard X-ray photons from the intense magnetar CXO 164710.2-455217, whose origin and spectral properties are discussed in Section \ref{sec:cxo}. The magnetar was excluded from the diffuse emission analysis within a circular region of 26.5 arcsec radius, consistent with the hard-to-soft brightness ratio profile that reaches its minimum at $\sim$24 arcsec (Fig.~\ref{fig:mag_radprof}), beyond which the scattered contribution becomes indistinguishable from the ambient diffuse emission. Region \#2, encompassing 4.45 arcminutes$^2$ (roughly three times larger than the previous region), accounts for the diffuse emission, with a total photon count of 8140. This region exhibits spectral hardening from the core outwards, indicating that stellar wind shocks are locally dominant but weaken at larger distances due to the diminishing massive star population. This is likely caused by strong wind-wind and/or wind-ISM interactions resulting from the massive stars concentrated in the core of the 
Wd\,1 cluster, a scenario further discussed in Section \ref{sec:emission}. The single-observation extraction methodology, including PSF modelling and point-source removal, is described in Appendix~\ref{sec:single}; detailed photon statistics per observation are tabulated in Appendix~\ref{ap:table}.
Although the HR map (Fig.~\ref{img_hr}) suggests the presence of additional spectral substructure within region~\#2, in particular a transitional softer zone between the cluster core and the magnetar-influenced emission to the west, which also differs in its underlying stellar population (see Fig.~\ref{fig:dif_mstars}), a further spatial subdivision is not feasible given the available photon statistics. The 8140 photons collected in region~\#2 across the full 1\,Ms dataset represent the minimum count required for a robust multi-temperature spectral fit with acceptably constrained model parameters. Subdividing this region into two independent apertures would reduce the photon count below this threshold, preventing any physically meaningful spectral characterisation of each sub-region independently. We therefore retain region~\#2 as a unified extraction aperture, while acknowledging that it likely encompasses physically heterogeneous emission that future, deeper observations could resolve.

Two distinct sets of areas are used in this work, each serving a different purpose. The morphological and hardness-ratio analysis uses the full projected extent of the diffuse emission above the background, down to the signal-to-noise limit of the flux and hardness maps (region~\#1: $1.47$~arcmin$^2$; region~\#2: $4.45$~arcmin$^2$); these areas characterise the spatial morphology and set the relative sizes of the two regions. The spectral analysis instead uses the extraction apertures listed in Table~\ref{tab:lumi} (region~\#1: $0.70$~arcmin$^2$; region~\#2: $7.07$~arcmin$^2$), defined by requiring the diffuse signal to exceed the background by at least twice the map signal-to-noise level, which minimises background contamination and yields more robust fits. Because the two sets optimise different goals, they do not need to coincide. The surface fluxes in Table~\ref{tab:lumi} are normalised by the extraction areas.

\subsection{Spectral modelling}

\begin{figure*}[ht!]
\includegraphics[width=9.3cm,angle=0]{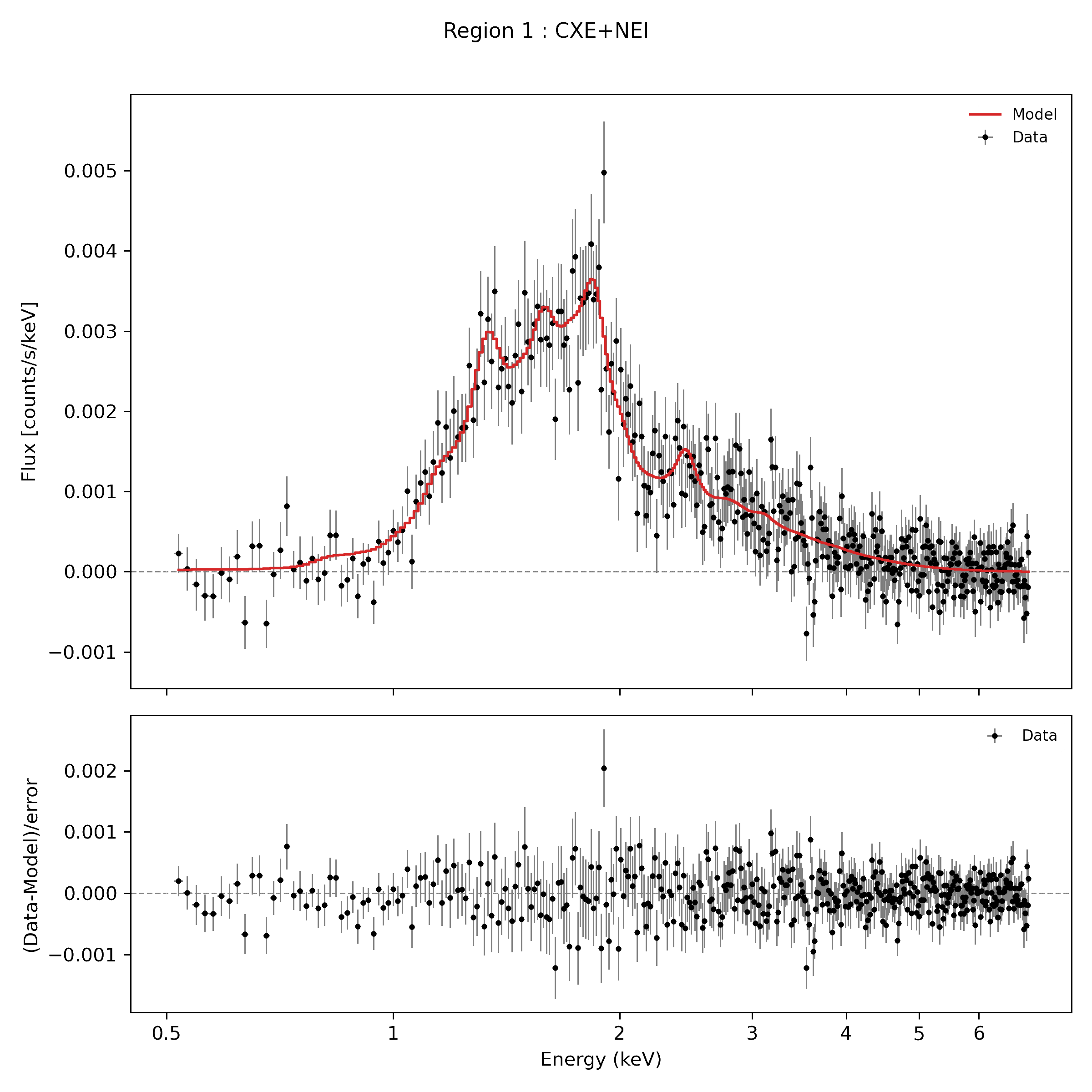}
\includegraphics[width=9.3cm,angle=0]{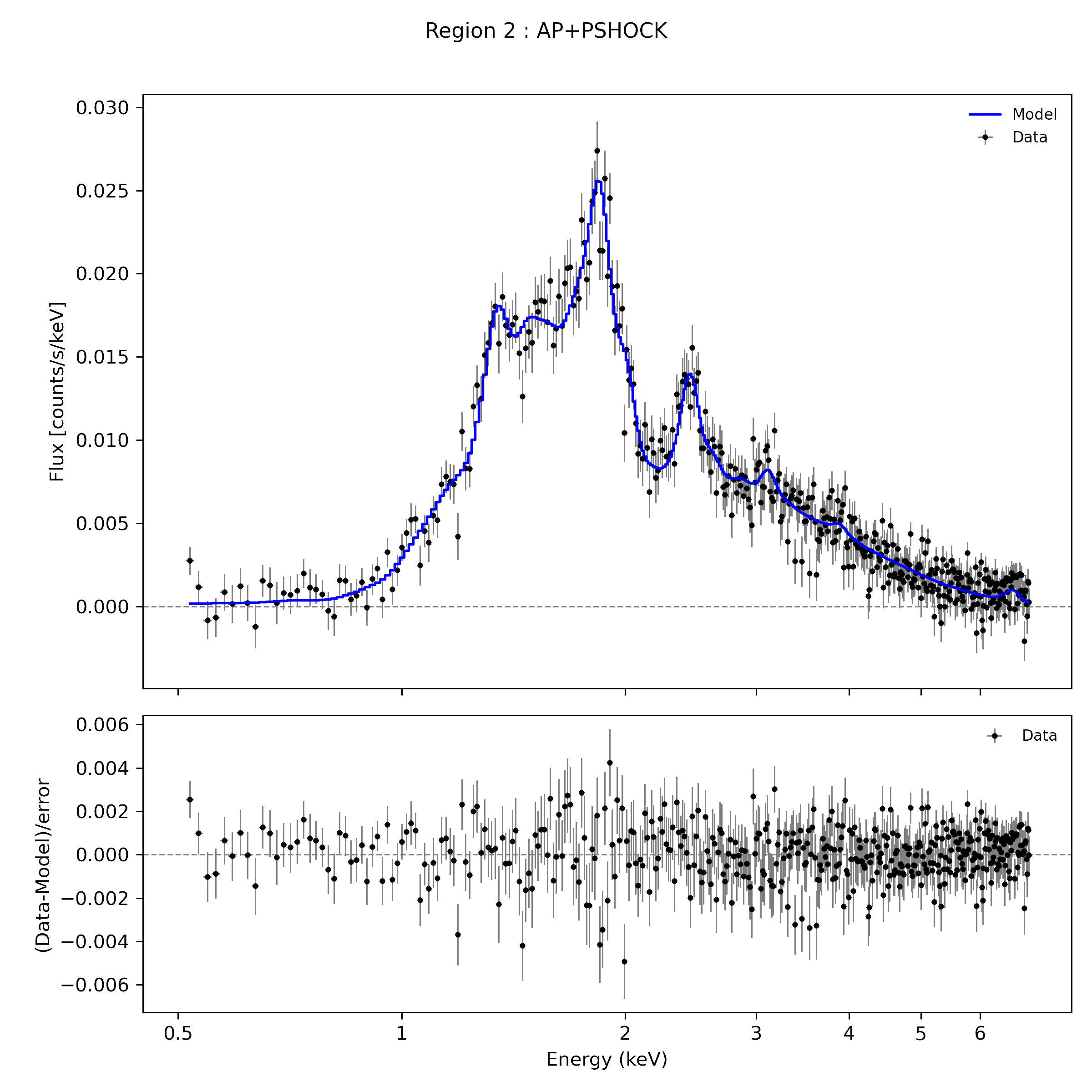}
\caption{\small Background-subtracted, stacked spectra of the diffuse X-ray emission of region~\#1 (left) and region~\#2 (right). In each panel the upper sub-panel shows the data (black points with error bars) and the best-fit adopted model: Region~\#1 is fitted with the CXE+NEI model (red line) and region~\#2 with the APEC+PSHOCK model (blue line). The lower sub-panel shows the fit residuals, (data$-$model)/error. The spectra are displayed in detector units, as the intrinsic model was convolved with the detector response matrices.}
\label{fig:img_spectra}
\end{figure*}

\begin{table*}[ht!]
\small
    \centering
    \caption{Spectral fit results for the diffuse emission of regions~\#1 and~\#2.}
    \begin{tabular}{lllllllllll}
\hline      
Model     &Residual&$N_\mathrm{H}$(10$^{22}$)&kT$_1$&Ab$_1$ &$\tau_1$(10$^{10}$)&N$_1$(10$^{-4}$)&kT$_2$&Ab$_2$ &$\tau_2$(10$^{10}$)&N$_2$(10$^{-4}$)\\
&C-stat/dof&[cm$^{-2}$]&[keV]& [Z$_\odot$]   &[s\,cm$^{-3}$]&[cm$^{-5}$]&[keV]& [Z$_\odot$]   &[s\,cm$^{-3}$]&[cm$^{-5}$] \\
\hline
Region 1\\
CXE+NEI $\dag$&418.3/438&1.84(f)&0.27$\pm$0.04&0.8(f)&$---$&0.66$\pm$0.47&1.09$\pm$0.05&0.14$\pm$0.03&7.7$\pm$3.6&5.39$\pm$0.56\\
NEI+NEI&417.9/435&1.72$\pm$0.26&0.66$\pm$0.55&0.47$\pm$0.49&10.9$\pm$29&2.97$\pm$4.9&1.24$\pm$0.23&0.06$\pm$0.09&3800(f)&3.42$\pm$2.0\\
NEI+AP&418.6/437&1.69$\pm$0.20&0.86$\pm$0.55&0.24$\pm$0.09&4.5$\pm$5.2&3.79$\pm$3.1&1.45$\pm$0.26&0.3(f)&$---$&1.49$\pm$0.95\\
\hline
Region 2\\
AP+AP&514.1/437&1.69$\pm$0.04&0.84$\pm$0.04&0.58$\pm$0.15&$---$&14.9$\pm$3.6&3.55$\pm$0.53&0.41$\pm$0.12&$---$&6.1$\pm$1.0\\
AP+NEI&453.7/436&1.66$\pm$0.04&2.37$\pm$0.23&2.1$\pm$2.6&$---$&2.3$\pm$2.7&2.32$\pm$0.28&0.36$\pm$0.09&2.5$\pm$0.4&7.9$\pm$2.2\\
AP+PS $\dag$&458.9/433&2.24$\pm$1.2&1.81$\pm$0.24&0.47$\pm$0.25&$---$&9.22$\pm$3.3&4.11$\pm$2.0&1.03$\pm$0.69&3.2$\pm$0.9&2.58$\pm$2.1\\
\hline
     \end{tabular}
    \tablefoot{Diffuse spectra of both regions are fitted with the Cash statistic (\texttt{cstat}); the goodness of fit is reported as the C-statistic value over the number of degrees of freedom (C-stat/dof). The model adopted for each region is marked with a $\dag$ symbol. PS refers to a PSHOCK model. For region~\#2 the adopted model uses an ionised absorber (\texttt{absori}); see Sect.~\ref{sec:spectral_region2}. All parameters correspond to XSPEC best-fit values, and the associated 1$\sigma$ confidence intervals were determined using the \texttt{error} command in XSPEC. Values marked with (f) indicate parameters that were frozen at the quoted value during spectral fitting.}
    \label{tab:spec}
\end{table*}

To extract the X-ray spectrum for each region, we used the \texttt{specextract} CIAO task on a diffuse event from each observation. All the spectra were properly weighted using calibration files such as ARFs and RMFs to ensure accuracy. The X-ray background spectra were obtained using stowed calibration event files. All X-ray spectra were grouped to achieve a minimum S/N of 1 per bin, ensuring unbiased best-fit values for the fitting procedure \citep{Albacete-Colombo2023a}. Because this grouping leaves of order one count per bin, the Gaussian assumption underlying the $\chi^2$ statistic is not satisfied; we therefore fit the diffuse spectra of both regions using the Cash statistic (\texttt{cstat} in XSPEC; \citealt{Cash1979}), which is based on the Poisson likelihood and provides unbiased parameter estimates in the low-count regime. The goodness of fit is assessed from the C-statistic value relative to the number of degrees of freedom and from the associated null-hypothesis probability. We retain the $\chi^2$ statistic only for the high-count fits, namely the magnetar point source (Sect.~\ref{sec:cxo}) and the radial surface-brightness profiles (Sect.~\ref{sec:single}), where the Gaussian regime applies.
Finally, we combined all individual spectra into a single spectrum using the CIAO task \texttt{combine\_spectra}. This task sums all extracted PHA spectra and combines them with the corresponding PHA background spectra and the ARF and RMF response files for the source and background.

Based on the hardness of the emission, we considered different combinations of emission models that are more realistic under the constraints of the photon statistics of the diffuse X-ray spectrum.
Spectral fitting was performed with a set of XSPEC spectral models \citep{Arnaud1996}. Of course, plasma emission is affected by the equivalent hydrogen absorption column ($N_\mathrm{H}$). To account for the effects of ISM absorption, we used the model \texttt{tbabs} \citep{Wilms2000}. This model considers a combination of $N_\mathrm{HI}$ (atomic hydrogen) and $N_\mathrm{H2}$ (molecular hydrogen). We allowed these values to change during the spectral fit to account for local ISM absorption fluctuations and the average absorption in the Wd\,1 direction, which accounts for the influence of stellar winds interacting with the ambient diffuse emission.

\begin{table*}[ht]
    \centering
    \caption{Surface fluxes and X-ray luminosities of the diffuse emission compared with previous studies.}
    \begin{tabular}{lcccccccc}
    \hline
 Results & Area  &$N_\mathrm{H}$ & kT &\multicolumn{2}{c}{Soft [0.5-2.0]} & \multicolumn{2}{c}{Hard [2.0-8.0]} \\
   from  & [arcmin$^2$] &(x10$^{22}$) & [keV] & f$_x^s$(x10$^{-13}$) & L$_x$(x10$^{33}$) & f$_x^s$(x10$^{-13}$)  & L$_x$(x10$^{33}$)\\
\hline
This work (4.2 kpc)   &&&&\\
Reg \#1$^{\star}$ & 0.70 &1.84&   0.27($\pm$0.04)& 0.43 & 0.06 & 
0.55  & 0.08 \\
&&&1.09($\pm$0.05)&&&\\
Reg \#2    & $\pi \times 1.5^2$ & 2.24 &1.81($\pm$0.24) &5.55& 8.29 &0.78&1.17  \\
&(=7.07)&&4.11($\pm$1.98)&&&\\
Total (reg1 + reg2)&7.77&&&5.09&
8.35&0.76&1.25\\
\hline
\cite{Muno2006} &&&&&&&\\
@5.0 kpc &$\pi$5$^2$ &2.2($\pm$0.3)& 0.7($\pm$0.3)& --- & --- & 1.2($\pm$0.3) & 30.0($\pm$0.9) \\
&(78.5)&&3.2($\pm$0.5)&&&\\
@4.2 kpc  &&&& --- & --- & 0.9($\pm$0.3) & 14.9($\pm$0.9) \\
\hline
\cite{Kavanagh2011} &&&&&&&\\
@3.6 kpc &$\pi$2$^2$& 2.03($\pm$0.13) & 0.68($\pm$0.12)&---  &---  &0.13($\pm$---)&2.56($\pm$--) \\
&(12.5)&&3.07($\pm$0.6)&&&\\
@4.2 kpc  &&&&---  &--- &0.13($\pm$--- )&3.48($\pm$--) \\
\hline
\cite{Haubner2025} &&&&&&&\\
@3.5 kpc  &$\pi$3.5$^2$ &2.2($\pm$0.2)&0.46($\pm$0.12)&---  &---    &1.0($\pm$0.5)&5.64($\pm$2.6) \\
&(38.5)&&5.7($\pm$2.5)&&&\\
@4.2 kpc  & &&&---  &---    &1.0($\pm$0.5)&8.12($\pm$2.6) \\

\hline
    \end{tabular}
    \tablefoot{X-ray surface fluxes $f_x^s$ [erg\,s$^{-1}$\,cm$^{-2}$\,arcmin$^{-2}$] and luminosities $L_x$ [erg\,s$^{-1}$] derived from the adopted spectral model of regions \#1 and \#2 at $d=4.2$~kpc, in the soft (0.5--2.0~keV) and hard (2.0--8.0~keV) bands. For region~\#2 we report absorption-corrected (intrinsic) quantities, since its spectral fit and absorbing column are well constrained; for region~\#1, we report observed (absorbed) quantities for the reason given in the footnote. Column~2 gives the areas of the spectral extraction regions. 
    For comparison entries, areas are given as $\pi r^2$ (in arcmin$^{2}$) where $r$ is the extraction radius in arcmin; the numerical value in parentheses is the equivalent area in arcmin$^{2}$. Surface fluxes $f_x^s$ are obtained by normalising the region flux by the corresponding extraction area; the Total row combines the observed region~\#1 and absorption-corrected region~\#2 quantities and is dominated by region~\#2. The hard-band fluxes and luminosities reported in columns~7 and~8 are compared with values from \cite{Muno2006}, \citet{Kavanagh2011}, and \cite{Haubner2025}, accounting for their adopted distances and extraction areas for Wd\,1. For consistency, these quantities were also recomputed assuming a distance of 4.2~kpc. We note, however, that these literature hard-band luminosities are biased upward by the inclusion of the bright magnetar CXO\,164710.2$-$455217, which dominates the hard X-ray output of Wd\,1 and which we explicitly excluded from our diffuse-emission analysis (Sect.~\ref{sec:cxo}); the earlier studies did not remove this contribution, so the comparison is not strictly one-to-one. }
    \label{tab:lumi}
\end{table*}

\subsubsection{Physical motivation for model selection}
\label{sec:modelmotivation}

The selection of spectral models is based on the physical conditions expected in each region, as inferred from the hardness ratio map (Section~\ref{sec_hr}) and the stellar population analysis (Fig.~\ref{fig:dif_mstars}). All models are implemented within the XSPEC framework \citep{Arnaud1996} and are convolved with a single absorbing component, \texttt{tbabs} \citep{Wilms2000}. We consider four plasma emission components, each associated with a distinct physical process, all extensively discussed in some young SFRs, such as the Carina Nebula \citep{Townsley2003b, Townsley2011a} and, more recently, in the Cygnus OB2 stellar association \citep{Albacete-Colombo2023a}:
 
\begin{itemize} 
\item APEC (AP): A thermal plasma in collisional ionisation equilibrium (CIE; \citealt{Smith2001}) is appropriate for hot gas where ionisation states have equilibrated with the electron temperature. This model serves as the canonical baseline for diffuse X-ray emission driven by stellar wind energy input in massive star-forming regions, where the thermalised plasma on parsec scales is well represented by one or two temperature components as the simplest CIE baseline.\\
 
\item NEI: A non-equilibrium ionisation model (\citealt{Borkowski2001}) parameterised by an ionisation timescale $\tau = n_e\,t$ that measures the departure from CIE. Appropriate wherever active wind-ISM shocks are occurring on timescales shorter than the ionisation equilibration time, as is generally expected in young, star-forming environments. NEI components have been required in deep spectral analyses of the diffuse emission in the Carina Nebula, where ongoing wind-ISM interactions prevent full ionisation equilibration.\\
 
\item PSHOCK (PS): A plane-parallel shock model that integrates over a continuous distribution of ionisation timescales from zero to $\tau_\mathrm{max}$, which is physically more appropriate than a single-timescale NEI model when shocks of many different ages coexist within the extraction aperture. This is the expected situation in the dense stellar core of Wd\,1, where dozens of massive stellar winds interact simultaneously. PSHOCK components have been specifically motivated in regions of high massive-star density, where simultaneous wind–wind collisions produce a continuous range of post-shock ionisation ages, as demonstrated.\\
 
\item CXE: Charge-exchange emission arises at the interface between fast, hot stellar-wind plasma and cold, neutral ISM material when highly charged ions capture electrons from the neutral donor into excited states. The captured electron is preferentially deposited into high principal quantum number ($n$) levels, and the subsequent radiative cascade to the ground state produces line emission across many series (not only K$_\alpha$, but also higher-$n$ transitions), with an enhanced high-$n$ line ratio that is the characteristic spectral signature of charge exchange relative to a collisionally ionised plasma at the same ionisation balance \citep{Liu2012}. We model this component with the AtomDB Charge eXchange model (ACX; \citealt{Smith2012, Smith2014}) as implemented in XSPEC. In ACX, the temperature parameter $kT_\mathrm{CXE}$ does not describe a Maxwellian electron population but sets the ion-population distribution of the recombining plasma as if it were in collisional ionisation equilibrium at that temperature; the emitted spectrum then arises purely from the charge-capture cascades onto that ion population. This process has been identified as a physically motivated component in deep X-ray observations of massive star-forming regions, where the boundary between hot stellar wind plasma and adjacent cold ISM structures mirrors the conditions expected in the periphery of Wd\,1.
\end{itemize}
 
Variable-abundance models (\textsc{vapec}, \textsc{vnei}) were not considered because the available photon statistics and CCD spectral resolution do not allow for meaningful constraints to be placed on individual element abundances; metallicities were instead varied within $Z = 0.3$--$2.0\,Z_\odot$ during fitting. For region~\#2, where the high stellar density and multiple simultaneous wind-wind interactions produce a broad range of shock ages, we
test AP+AP as the CIE baseline, NEI+NEI for non-equilibrium corrections, and AP+PS as the physically preferred model , following the precedent established in the Carina Nebula core \citep{Townsley2011a} and Cygnus~OB2 \citep{Albacete-Colombo2023a} where PSHOCK components consistently improved spectral fits over single-timescale NEI in regions of high wind-interaction rate. For region~\#1, where the reduced wind pressure
and proximity to denser ISM structures suggest the presence of charge-exchange emission, we additionally tested the CXE+NEI combination against models where the soft emission is reproduced by collisional processes alone (CIE and non-equilibrium ionisation), motivated by the CXE detections at the wind--ISM interfaces in the periphery of Carina \citep{Townsley2011a} and Cygnus~OB2 \citep{Albacete-Colombo2023a}.

\subsubsection{Region 1: Outer soft-emission region}

Region~\#1 lies at the cluster periphery where the collective wind pressure is significantly reduced, and the hot gas is in direct contact with denser, cooler ISM structures. The purely collisional models provide acceptable but imperfect descriptions: NEI+NEI (C-stat/dof $=417.9/435$) and NEI+AP (C-stat/dof $=418.6/437$) reproduce the overall continuum but force the second thermal component to $kT \simeq 1.2$--$1.5$~keV while leaving the softest emission below $\sim$1~keV under-modelled. The adopted CXE+NEI model (C-stat/dof $=418.3/438$; null-hypothesis probability $0.82$) is statistically comparable and is preferred because it explicitly accounts for the soft excess through charge-exchange emission at the hot/cold gas boundary, while the NEI component describes the recently shocked hot outflow. The best-fit parameters are $kT_\mathrm{CXE} = 0.27 \pm 0.04$~keV, $kT_\mathrm{NEI} = 1.09 \pm 0.05$~keV, and $N_\mathrm{H} = 1.84 \times 10^{22}$~cm$^{-2}$, with a short NEI ionisation timescale $\tau \simeq 7.7 \times 10^{10}$~s~cm$^{-3}$ consistent with recent wind--ISM interactions \citep{Smith2010a}. We stress that $kT_\mathrm{CXE}$ is not the temperature of a second emitting plasma but the ACX parameter that fixes the ionisation (charge-state) balance of the recombining ions; a coexisting plasma at $\sim$0.27~keV is therefore not required for the charge-exchange emission to arise. The fact that this ionisation balance ($kT_\mathrm{CXE} = 0.27$~keV) lies well below the NEI electron temperature ($kT_\mathrm{NEI} = 1.09$~keV) is precisely what is expected for the strongly under-ionised plasma indicated by the short NEI ionisation timescale ($\tau \simeq 7.7 \times 10^{10}$~s~cm$^{-3}$, well below the $\tau \gtrsim 10^{12}$~s~cm$^{-3}$ required to reach collisional ionisation equilibrium): the ion charge states lag the electron temperature, so the ion population that undergoes charge capture corresponds to a lower effective ionisation temperature than that of the freshly shocked gas. The two values are thus consistent descriptions of a single, recently shocked, under-ionised outflow rather than two components at incompatible temperatures.

No signs of the FeK$_\alpha$ complex were detected in region~\#1, consistent with the weaker wind-ISM interactions and lower plasma temperatures in this peripheral region. The observed (absorbed) diffuse X-ray luminosity of region~\#1 is $1.5\times10^{32}$~erg~s$^{-1}$ in the 0.5--8.0~keV band, split roughly evenly between the soft ($6.4\times10^{31}$~erg~s$^{-1}$) and hard ($8.2\times10^{31}$~erg~s$^{-1}$) bands. The report observed (absorbed) luminosities is dominated by a very soft component ($kT_\mathrm{CXE}\simeq0.27$~keV) seen through a substantial column ($N_\mathrm{H}\simeq1.84\times10^{22}$~cm$^{-2}$), so the corresponding absorption correction in the soft band becomes extreme ($\gtrsim10^{2}$) and, thus, the absorption-corrected (intrinsic) luminosity is an ill-constrained upper limit.

\subsubsection{Region 2: Cluster core}
\label{sec:spectral_region2}

Among the tested models for region~\#2, the double-APEC model (AP+AP) provides the poorest description (C-stat/dof $=514.1/437$; null-hypothesis probability $1.8\times10^{-3}$), with the two thermal components settling at $kT \simeq 0.84$ and $3.55$~keV. The APEC+NEI model yields a markedly better fit (C-stat/dof $=453.7/436$; null-hypothesis probability $9.5\times10^{-2}$) with two comparable temperatures ($kT \simeq 2.37$ and $2.32$~keV). The adopted APEC+PSHOCK model (AP+PS; C-stat/dof $=458.9/433$; null-hypothesis probability $5.4\times10^{-2}$) is statistically comparable to APEC+NEI and is preferred on physical grounds: the PSHOCK component reproduces the continuous range of post-shock ionisation timescales (with an upper limit $\tau_u \simeq 3.2 \times 10^{10}$~s~cm$^{-3}$) generated by the many coexisting stellar-wind shocks in the dense core, which a single-timescale NEI cannot capture. The best-fit parameters of the adopted model are $kT_\mathrm{PS} = 4.11 \pm 1.98$~keV and $kT_\mathrm{AP} = 1.81 \pm 0.24$~keV, with an absorbing column $N_\mathrm{H} = 2.24 \pm 1.20 \times 10^{22}$~cm$^{-2}$. For this region the absorption is modelled with an ionised absorber (\texttt{absori}) rather than the neutral \texttt{tbabs} adopted elsewhere: the intense UV radiation field of the massive stars in the cluster core photoionises the surrounding neutral hydrogen, and because the photoelectric absorption cross-section of partially ionised gas differs from that of neutral hydrogen, an ionised-absorber treatment provides a more physically appropriate description of the absorbing medium local to Wd\,1.

Further evidence of the harder, shock-dominated emission in region~\#2 is the detection of the Fe\,K complex at 6.4--6.7~keV. Plasma temperatures sufficient to produce significant He-like Fe\,\textsc{xxv} emission exceed $\sim$2.0~keV \citep{Raassen2003} and are typically associated with wind--wind or wind--ISM shocks \citep{Pittard2010}. Recent X-ray spectral analyses of WR stars in Wd\,1 confirm the presence of the Fe\,\textsc{xxv} line at $\sim$6.7~keV; the fluorescent Fe\,\textsc{i} line at $\sim$6.4~keV has been detected in several sources, indicating the coexistence of dense, cold material with shocked hot plasma \citep{Anastasopoulou2024}. The absorption-corrected diffuse X-ray luminosity of region~\#2 is $9.5\times10^{33}$~erg~s$^{-1}$ in the 0.5--8.0~keV band, with $\sim$88\% in the soft band ($8.3\times10^{33}$~erg~s$^{-1}$) and $\sim$12\% in the hard band ($1.2\times10^{33}$~erg~s$^{-1}$), consistent with the collective interaction of multiple stellar winds driving a hot, multi-temperature plasma \citep{Canto2000}.

\subsubsection{Spectral results and luminosity comparison}

In Figure~\ref{fig:img_spectra} we present the stacked X-ray spectra of both regions; Table~\ref{tab:spec} lists the best-fit parameters for all tested models, with the adopted model for each region marked with a $\dag$ symbol. The most notable result is the pronounced temperature contrast: Region~\#2 exhibits a significantly hotter plasma ($kT \sim 1.81$--$4.11$~keV) than region~\#1 ($kT \sim 0.27$--$1.09$~keV), and the Fe\,K detection in region~\#2 absent in region~\#1 independently supports a physically distinct origin for the two emission components. The discussion of the absorption column densities and their comparison with previous studies is presented in Section~\ref{discussion}.

The X-ray luminosities quoted below were derived from the adopted model ($\dag$) of each region using the \texttt{flux} command in XSPEC over the 0.5--8.0~keV band, assuming a source distance of $d = 4.2$~kpc ($4\pi d^2 = 2.11\times10^{45}$~cm$^2$). From the adopted spectral models, the observed (absorbed) diffuse X-ray luminosity of region~\#1 is $1.5\times10^{32}$~erg~s$^{-1}$, while the absorption-corrected luminosity of region~\#2 is $9.5\times10^{33}$~erg~s$^{-1}$ (0.5--8.0~keV; Table~\ref{tab:lumi}), so that the total diffuse luminosity, dominated by the cluster core, is $\simeq9.6\times10^{33}$~erg~s$^{-1}$. The choice of observed versus absorption-corrected quantities is deliberate and differs between the two regions. The diffuse emission of region~\#1 is dominated by a very soft component (the charge-exchange ACX term, $kT_\mathrm{CXE}\simeq0.27$~keV) seen through a substantial absorbing column ($N_\mathrm{H}\simeq1.84\times10^{22}$~cm$^{-2}$). For such a cold, heavily absorbed plasma the bulk of the intrinsic emission is radiated below the observable band, so recovering the unabsorbed flux requires extrapolating the model far outside the energy range actually constrained by the data. This makes the absorption correction in the soft band extreme (a factor $\gtrsim10^{2}$) and intrinsically uncertain: within the $1\sigma$ parameter ranges the corrected soft flux varies by about a factor of two; namely, its lower bound already differs from the best-fit value by a comparable amount, and the inferred intrinsic luminosity becomes essentially unbounded from above, formally exceeding the integrated luminosity of the whole cluster. Such a large model extrapolation could jeopardise the physical interpretation, so the absorption-corrected luminosity of region~\#1 must be regarded as an ill-constrained upper limit, and we report its observed (directly measured) value instead. For region~\#2, by contrast, the spectral fit and the absorbing column are well constrained and the emission is substantially harder, so the absorption correction is moderate and the absorption-corrected (intrinsic) luminosity is reliable; we therefore quote it in the standard way.

For the wind-to-X-ray efficiency analysis described in Section~\ref{sec:emission}, we adopted the region~\#2 luminosity, as it traces the cluster-wind-driven emission from the massive stellar core; region~\#1 is dominated by softer CXE processes at the cluster periphery.

Table~\ref{tab:lumi} presents a direct comparison with earlier works. The high hard-band luminosity reported by \citet{Muno2006} ($\sim2.1\times10^{34}$~erg~s$^{-1}$, over 78.5~arcmin$^2$) reflects the much larger integration volume and the limited point-source subtraction achievable with two short observations. The more conservative analysis of \citet{Kavanagh2011} ($0.25\times10^{34}$~erg~s$^{-1}$, 12.5~arcmin$^2$) and our estimate ($0.12\times10^{34}$~erg~s$^{-1}$, $\sim$7.3~arcmin$^2$) are in better agreement. \citet{Haubner2025} report $0.56\times10^{34}$~erg~s$^{-1}$ from 38.5~arcmin$^2$. An important caveat is that all of these literature hard-band luminosities are biased upward by the inclusion of the bright magnetar CXO\,164710.2$-$455217, the dominant hard X-ray source in Wd\,1, which the earlier studies did not subtract; in our analysis the magnetar and its scattered halo are explicitly excluded (Sect.~\ref{sec:cxo}), so our lower hard-band luminosity partly reflects this cleaner discrimination of the diffuse component. These systematic differences, discussed further in Section~\ref{discussion}, reflect the differing choices of extraction area and point-source subtraction methods across studies.

\begin{figure*}[h!]
\includegraphics[width=18.3cm,angle=0]{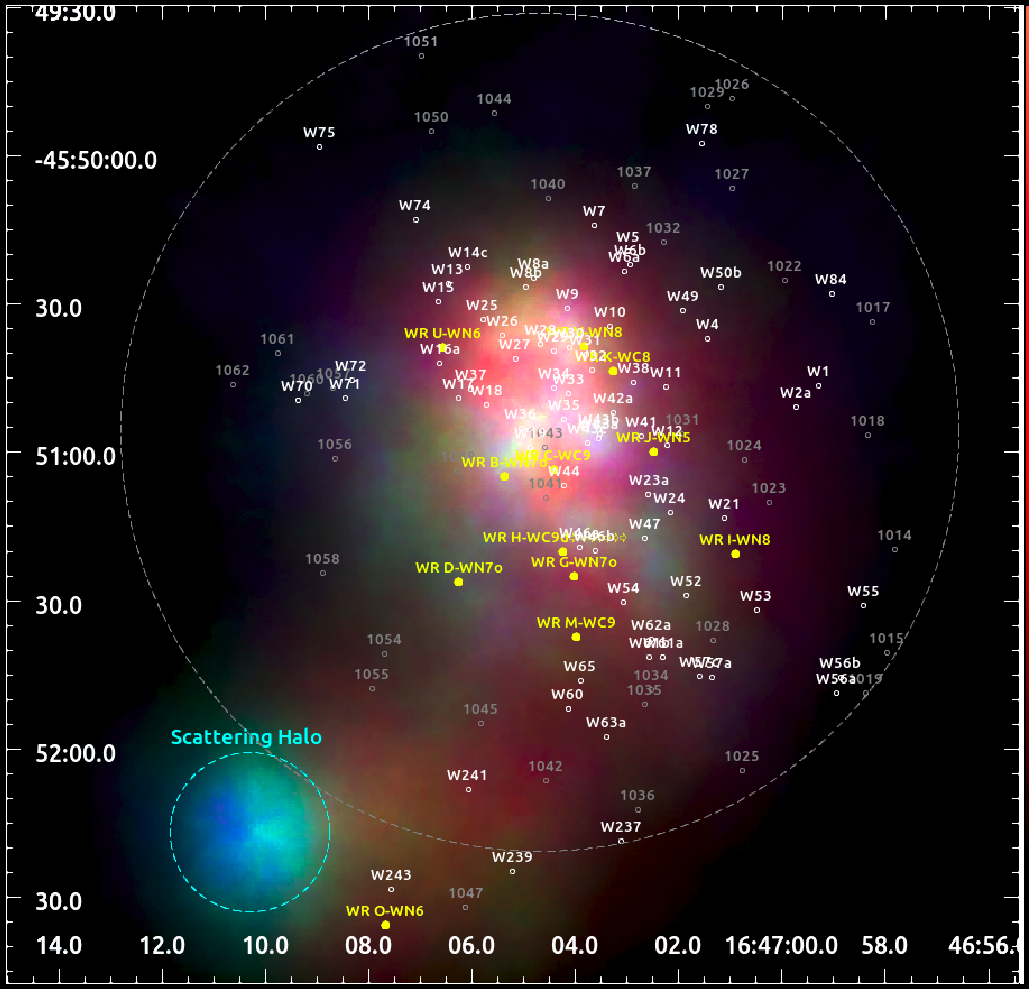}
\caption{\small Flux-intensity map (ph\,s$^{-1}$\,cm$^{-2}$) in linear scale computed in the soft (0.5–1.2~keV; red), medium (1.2–1.9~keV; green), and hard (1.9–7.0~keV; blue) energy bands. The large dashed grey circle indicates the computed radius of the cluster, while the dashed cyan circle highlights the region influenced by scattered X-ray photons from the magnetar CXO~164710.2$-$455217. White labels refer to known WR stars, whereas grey-coloured identification numbers indicate 72 evolved O-type massive stars (classes III, II, and I), as reported by \citet{Clark2022}.
Yellow-filled circles indicate the most common names for the WR stars in Wd1 \citep{Anastasopoulou2024}.}
\label{fig:dif_mstars}
\end{figure*}

\section{Nature of the emission}
\label{sec:emission}

While the south-eastern hard diffuse X-ray emission is likely associated with scattered radiation from the magnetar CXO~164710.2$-$455217, we cannot rule out the presence of a combination of thermal and non-thermal components coexisting within the same region. Figure~\ref{fig:dif_mstars} shows that hard X-ray emission dominates the central core of the cluster, whereas a more extended intermediate and soft X-ray component is detected in the outer regions. In the following sections, we discuss both thermal and non-thermal processes and the astrophysical conditions required for their presence.

\subsection{Non-thermal emission}

Non-thermal (NT) processes can also produce diffuse X-ray emission, primarily through synchrotron losses (SLs) and inverse Compton (IC) scattering. Both mechanisms generate hard X-rays ($E > 2$~keV), though on different spatial scales. There is marginal evidence for NT activity in this region \citep{Muno2006}, suggesting that NT particles may be present, possibly accelerated by colliding stellar winds \citep[e.g.][]{Eichler1993}. Such high-energy particles interact with the surrounding gas or radiation fields and may produce observable $\gamma$-ray emission through hadronic ($pp \rightarrow \pi^0 \rightarrow 2\gamma$) and leptonic (IC or Bremsstrahlung) channels. The spatial and energetic correlation between the diffuse X-ray and $\gamma$-ray components supports that both originate from the same population of accelerated particles within Wd\,1 \citep{HESS_Wd1_2022, Lemoine-Goumard2024}.

The above argument is based on the assumption that electrons are accelerated within a colliding-winds environment of size $\gtrsim 0.1$ pc via scattering off magnetic fluctuations close downstream to stellar termination shocks and lose energy therein due to the short cooling time, as discussed below. 
If electrons stochastically accelerate via a second-order process within the colliding stellar winds, the acceleration time scale can be estimated as $T_\mathrm{acc} \sim  (L/c)(V/c)^{-2}$, where $L$ is the size of the accelerating region and $V$ is an average stellar winds speed at that location. For $L \sim 0.1$ pc and $V \sim 10^3$ km/s, we find $T_\mathrm{acc} \sim 10^3 - 10^4$ years.
Although $T_\mathrm{acc}$ is short compared to the cluster lifetime, it is longer than the loss time scale via IC (UV-driven). In the Thomson limit ($4\varepsilon_{UV} \gamma/ m_e c^2 \ll 1$, where $\gamma$ is the accelerated electrons Lorentz factor and $\varepsilon_{UV} = 5$ eV is the UV initial photon energy), the IC loss time scale at a distance, $R$, from the stellar surface \citep{Fraschetti.etal:23} is expressed as
\begin{equation}
t^T_{IC} (\gamma) = \frac{m_e c^2 \gamma}{P^T_{IC} } \, \left(\frac{R}{R_{\star}}\right)^2 \nonumber\\ 
= \frac{ 1.1 \times 10^7 \, {\rm yr}}{\gamma } \left( \frac{3.5 \times 10^4\, K}{{ T_{\star}}} \right)^4 \left(\frac{R}{0.1 \, \rm{pc}}\right)^2 ,
\label{t_loss_Thom}
\end{equation}
where we have used for a WR-type a stellar surface temperature, $ T_{\star} \sim 35,000$ K, and radius, $R_{\star} \sim 20 \, R_\odot$, along with $U_\mathrm{rad} = \sigma T_{\star}^4 /c$, where $\sigma$ is the Stefan-Boltzmann constant, and the factor $({R}/R_{\star})^2$ accounts for the scaling of $U_\mathrm{rad}$. Thus, for multi-GeV electrons ($\gamma \sim 10^4$), the cooling is short enough ($t^T_{IC} < T_\mathrm{acc}$) to lead to efficient IC losses close downstream to the stellar termination shock. Similar conclusions can be drawn for different star types (e.g. O-star).

To assess the relative importance of SLs and IC losses, we calculated the ratio between the radiation field energy density ($U_{\rm ph}$) produced by the massive stars in the region and the expected magnetic energy density of the ISM ($U_{B} = B^2 / 8\pi$). 
We adopted a typical ISM magnetic field strength of $B \sim 2~\mu$G, following the approximations provided by \cite{Harer2023}. The individual $U_{B}/U_{\rm ph}$ ratio was computed for both single main-sequence and evolved massive stars. To evaluate which NT process is likely to dominate in the region, we estimated the projected 2D and 3D surface density of massive stars in the core of Wd\,1. 

To estimate the 3D stellar density within the cluster, we consider the circular extraction area of a radius, $R = 1.5$~pc. Assuming a spherical geometry, the total volume is $V_{\text{sph}} = \frac{4}{3}\pi R^3 \approx 14.1~\text{pc}^3$, which yields a volumetric density of $n_{\text{massive}} \approx 12.7~\text{stars~pc}^{-3}$. Alternatively, adopting a cylindrical geometry with a characteristic depth equal to the cluster diameter ($L = 2R = 3$~pc) to account for the line-of-sight integration, the volume is $V_{\text{cyl}} = \pi R^2 L \approx 21.2~\text{pc}^3$, resulting in a density of $n_{\text{massive}} \approx 8.5~\text{stars~pc}^{-3}$. These values provide a representative range for the spatial concentration of the $\sim 180$ massive stars reported in the literature \citep{Clark2005, Crowther2007, Negueruela2010}, consistent with the calculated surface density of $\Sigma_{\text{massive}} \approx 25~\text{stars~pc}^{-2}$ over the projected area of $7.1~\text{pc}^2$.

Based on the population of about $\sim$50 O-type stars, $\sim$80 OB supergiants, $\sim$24 WR stars, 9 red supergiants (RSGs), and at least 6 luminous blue variables (LBVs) or yellow hypergiants (YHGs), we estimated the total bolometric luminosity by adopting representative average luminosities for each subtype: 5\,$\times$\,10$^5$\,L$_\odot$ for O-type stars, 2\,$\times$\,10$^5$\,L$_\odot$ for OB supergiants, 7\,$\times$\,10$^5$\,L$_\odot$ for WR stars, 1\,$\times$\,10$^5$\,L$_\odot$ for RSGs, and 7\,$\times$\,10$^5$\,L$_\odot$ for LBVs/YHGs. Thus, we obtain a total bolometric luminosity of approximately 7\,$\times$\,10$^7$\,L$_\odot$ ($\sim$2.7\,$\times$\,10$^{41}$\,erg\,s$^{-1}$), or 70 L$_6$, where $L_6$ is the total stellar luminosity computed in $10^6~L_\odot$ solar luminosities. Considering the collective contribution of all evolved massive stars in their vicinity, the ratio of synchrotron to inverse Compton (IC) losses is given by
\[
\frac{L_{\mathrm{syn}}}{L_{\mathrm{IC}}} = \frac{U_B}{U_{\mathrm{ph}}} \approx 7.1 \times 10^{-4} \frac{B^2}{L_6} \approx 0.03.
\]

This implies that IC losses dominate over SLs in the immediate vicinity of evolved massive stars, with UV-driven IC scattering occurring primarily within the inner zones of stellar winds at distances $\lesssim 0.1$ pc from the stellar sources. Consequently, relativistic electrons accelerated in the dense cluster core are unlikely to propagate far from their sites of acceleration without severe energy losses, and large-scale ($\gtrsim 0.1$ pc) IC cooling within the core is not expected to contribute significantly to the diffuse X-ray emission observed in the region. However, these cooling arguments apply specifically to the dense inner regions of the cluster core, where the radiation field energy density is extremely high. However, it should not be interpreted as contradicting the established GeV and multi-TeV $\gamma$-ray emission detected from much larger scales around Wd\,1 \citep{HESS_Wd1_2022, Lemoine-Goumard2024}. The H.E.S.S. and \textit{Fermi}-LAT instruments have spatial resolutions of $\sim$0.1$^{\circ}$ ($\sim$7~pc) and larger, respectively. This is much coarser than the $\sim$2~pc core region and therefore cannot resolve the inner cluster physics; instead, it detects emission from a volume extending $\sim$10--100~pc, where accelerated particles likely escape from the dense core into regions with significantly diluted radiation fields. This allows them to reach energies of at least 100~TeV before producing observable $\gamma$-ray emission via IC or hadronic processes.

Thus, while local IC cooling efficiently suppresses non-thermal X-ray emission, making negligible the contribution of non-thermal mechanisms to the diffuse X-ray luminosity at the core of Wd\,1, it seems to be a highly efficient cosmic ray accelerator on larger scales. So, the extended GeV and multi-TeV $\gamma$-ray emission detected from the surrounding region is powered by particles that successfully escape the radiation-dominated core and reach energies of at least 100~TeV in the diluted ISM at distances of $\sim$10--100~pc.

We also investigated whether a component relative to synchrotron radiation from TeV electrons was detectable in our dataset. Following the extraction methodology described in Appendix~\ref{sec:single}, we extracted spectra from the outskirts of Wd\,1, where the shocks of evolved stars are expected to collide and where the thermal emission is the faintest. However, we did not find statistically significant improvements in the fit statistics when including a power-law component in the spectra. We were only able to estimate a $3\sigma$ flux upper limit by considering a power-law component with $\Gamma$ fixed at 2.5 of $1.6\times10^{-14}$~erg~cm$^{-2}$~s$^{-1}$ in the 4--6~keV range and $5.8\times10^{-14}$~erg~cm$^{-2}$~s$^{-1}$ in the 0.5--8~keV range. These upper limits are consistent with the expectation that IC losses efficiently suppress synchrotron emission in the radiation-dominated environment of the cluster core, as discussed above.

\subsection{Thermal emission}
We can estimate the diffuse X-ray luminosity using an advanced analytical approach that examines the physical mechanisms underlying the wind-ISM interaction. It has been solved analytically in the framework of the Cluster Wind Model (CWM) \citep{Canto2000}. Such a model assumes that the bulk of the diffuse emission in the region is due to a hot plasma that exhibits relaxed, centre-filled morphology with a lack of any obvious, measurable temperature gradients. Therefore, we considered the simple hypothesis of a uniform, optically thin thermal plasma with a simple geometry, although the emission is undoubtedly more complex. The winds of these massive stars produce a hot, low-density gas that, under certain conditions, emits X-rays that are scattered in the environment. This process effectively heats the ISM and contributes to the expansion of the heated gas on the order of parsecs \citep{Lancaster2021b}. This influence extends beyond regions directly populated by massive stars. Even in areas where massive stars are not dominant, the interaction of stellar winds with the ISM induces significant changes in both its temperature and structure. 

Stellar winds provide a first-order estimate of the volumetric hydrogen density ($n_0$) and plasma temperature ($T_0$) of the intracluster medium (ICM), under the assumption that the stars are uniformly distributed within the outer radius, $R_c$, of the cluster. The collective mass and momentum injection from stellar winds thermalise through wind–wind shocks, producing the hot X-ray emitting gas. We use Eqs. 33 and 34 of \citet{Canto2000}. Following the analytical expressions, we obtained
\begin{equation}
    n_0 [{\rm cm^{-3}}] = 0.1\,N\,\dot M\,V_w^{-1} R_c^{-2}, \hskip 1cm 
    T_0 [{\rm K}] = 1.55\times10^7\,V_w^2,
\end{equation}
where $N$ is the number of contributing stars, $\dot{M}$ [10$^{-5}$ M$_\odot$ yr$^{-1}$] is the mean mass-loss rate, $V_w$ [10$^3$ km s$^{-1}$] is the average wind velocity, and $R_c$ [pc] is the cluster radius containing the diffuse emission. 

To estimate the contribution from stellar winds, we adopt two approaches. Firstly, we estimate $n_0$ and $T_0$ from population synthesis using pySTARBURST99 \citep{Hawcroft2025}, following a similar approach to \citet{Larkin2025YMC}. This has the advantage of accounting for mass-loss from Cool Supergiant (CSG) stars, which can mass-load and, thus, decelerate a cluster's collective wind \citep{Larkin2025RSG}. We assume a single population with an initial mass of $5\times10^4\mathrm{M_{\odot}}$ and Milky Way metallicity, with all other parameters set to default values as in \citet{Larkin2025YMC}. We note that this does not account for supernovae. We show the stellar wind luminosity, $L_{\text{wind}}$, average cluster wind velocity, $\Bar{V}_{\text{cl}}$, and total stellar wind mass loss rate from the cluster, $\dot{M}_{\text{cl}}$, as a function of cluster age in Fig. \ref{fig:fig7}: 

\begin{figure}
    \centering
    \includegraphics[width=7cm,angle=0]{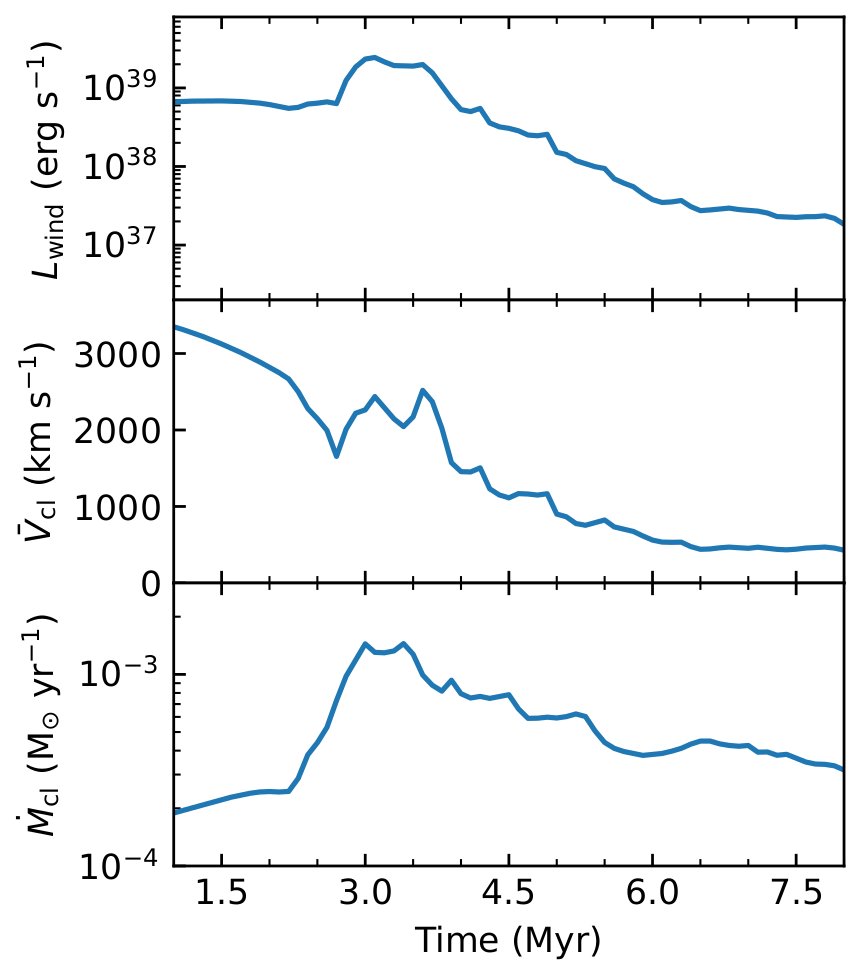}
    \caption{Cluster wind properties from pySTARBURST99: Stellar wind luminosity, average cluster wind velocity, and total wind mass-loss rate for a single-population cluster of an initial mass of $5\times10^4~\mathrm{M_{\odot}}$.}
    \label{fig:fig7}
\end{figure}

Assuming a single stellar population and age of 6~Myr, this model yields $\dot{M}_{\text{cl}}$ $\sim3\times10^{-4}~\mathrm{M_{\odot}~yr^{-1}}$ and $\Bar{V}_{\text{cl}}$ $\sim1000~\mathrm{km~s^{-1}}$. Given the sensitivity of $\Bar{V}_{\text{cl}}$ to uncertainties in CSG mass-loss rates \citep{Larkin2025YMC}, we adopted a range of $1000-1500~\mathrm{km~s^{-1}}$ for $\Bar{V}_{\text{cl}}$. This approach is equivalent to that of \citet{Canto2000}, where instead of defining average quantities per star and multiplying by the total number of stars, we already accounted for the effects of different spectral types. Taking $R_c = 1.5$ pc, $N\dot{M} = \dot{M}_{\text{cl}}$ and $V_w = \Bar{V}_{\text{cl}}$, this yields $n_0 \simeq 0.89-1.3~\text{cm}^{-3}$ and $T_0 \simeq1.6-3.5\times10^7~\text{K}$.

Our second approach is to use the measured wind parameters for the 24 known WR stars in the cluster to approximate $L_{\text{wind}}$, as WR stars likely dominate $L_{\text{wind}}$ in a YMC when active \citep{Larkin2025YMC}. For the 22 stars included in \citet{Rosslowe2015}, we used their wind values. For WR L, we took the average values for this spectral type from \citet{SANDER2019} and for WR S, we took the values measured by \citet{Fenech2018}. This yields $L_{\text{wind}}\sim4\times10^{38}~\text{erg s}^{-1}$.

Adopting $R_c=1.5$~pc, $N=24$, $\dot{M}=0.65\times10^{-5}~\mathrm{M}_\odot~\mathrm{yr}^{-1}$ (indicating a mass-loss rate of the 24 WR stars in Wd\,1, derived from the compilation of \citealt{Rosslowe2015}, \citealt{SANDER2019}, and \citealt{Fenech2018}), and $V_w=1.7\times10^3$~km~s$^{-1}$ (luminosity-weighted mean terminal wind velocity of the same 24 WR stars, derived from the compilation described above), we derived $n_0\simeq0.41$~cm$^{-3}$ and $T_0\simeq4.5\times10^7$ K. The emitting volume depends on the assumed geometry. For a spherical distribution, $V_{\rm sph}=\tfrac{4}{3}\pi R_c^3 \simeq 4.2\times10^{56}$~cm$^3$, while for a cylindrical geometry with depth $L=2R_c=3$~pc, $V_{\rm cyl}=\pi R_c^2 L \simeq 6.2\times10^{56}$~cm$^3$. The cylindrical case does not arise from a physical model specific to Wd\,1, but it serves as a geometric upper bracket on the emitting volume. X-ray observations constrain only the projected 2D morphology on the sky; the line-of-sight depth of the diffuse emission is observationally unconstrained. Setting $L = 2R_c$ (depth equal to the projected diameter) is therefore a conservative upper limit under the assumption that the emitting region is not more extended along the line of sight than across the plane of the sky --- consistent with the analogous treatment applied to the stellar volume density in Section~\ref{sec:emission}. We note that the \citet{Canto2000} cluster wind model formally assumes spherical symmetry, so the spherical estimate ($V_{\rm sph}$) is the physically self-consistent one; the cylindrical case is included solely to quantify the sensitivity of the result to the assumed geometry, and accounts for approximately half of the quoted $\sim$50\% uncertainty on $L_X^{\rm CWM}$. The corresponding emission measures are $EM_{\rm sph}=V_{\rm sph}\,n_0^2 \simeq 7.0\times10^{55}$~cm$^{-3}$ and $EM_{\rm cyl}=V_{\rm cyl}\,n_0^2 \simeq 1.0\times10^{56}$~cm$^{-3}$, which yield XSPEC normalisations of $3.3\times10^{-4}$ and $5.0\times10^{-4}$~cm$^{-5}$ for a distance of 4.2~kpc, respectively. These values are within a factor of $\sim$2 of the spectral fits reported for region~\#2 (Table~\ref{tab:spec}), where we measured $\sim1.0\times10^{-3}$~cm$^{-5}$. The stacked spectral products used for this comparison were constructed from the per-observation extraction described in Appendix~\ref{ap:analysis}; the underlying photon counts per observation are listed in Appendix~\ref{ap:table} (Table~\ref{tab:tab_obs}).

Using the adopted APEC+PSHOCK (AP+PS) spectral model for region~\#2, with best-fit temperatures $kT_1 = 1.81$~keV and $kT_2 = 4.11$~keV and absorption column $N_{\rm H} = 2.24\times10^{22}$~cm$^{-2}$ (Table~\ref{tab:spec}), the predicted diffuse cluster-wind luminosity in the 0.5--8.0~keV band is $L_X^{\rm CWM}\simeq2.1\,(\pm1.0)\times10^{34}$~erg~s$^{-1}$, where the estimated $\sim$50\% uncertainty reflects the range of geometric assumptions (spherical vs.\ cylindrical emitting volume) and the adopted wind parameters. This value exceeds our observational estimate by a factor of approximately 2.2, indicating that the absorption-corrected luminosity derived from the spectral modelling does not fully agree with the predictions from the cluster wind formalism. This discrepancy can be explained by the low density of the ISM, which is caused by strong stellar winds sweeping out the interior of the cluster. As a result, the efficiency of converting the mechanical energy of the stellar wind into thermal X-ray emission is significantly reduced \citep[e.g.,][]{Stevens2003, Townsley2003}. Given that $L_X \propto n^2$, the observed factor of $\sim$2.2 discrepancy between the CWM prediction and the measured luminosity implies an effective gas density reduction of\,$\sim\!\sqrt{2.2} \approx 1.5$ relative to the idealised uniform-density assumption. This is consistent with the anti-correlation between infrared emission and diffuse X-rays discussed in Section~\ref{discussion}, which independently supports a partially evacuated ICM. Therefore, the observed luminosity reflects a more dilute plasma environment than the idealised predictions of the model, highlighting the role of radiation-driven ISM evacuation in shaping the X-ray properties of diffuse emission in massive, dense star clusters.

\section{Discussion}
\label{discussion}

Wd\,1 emerges as a unique example of a young, massive starburst cluster, not only due to its dense population of massive stars, but also because of its distinctive diffuse X-ray morphology and properties. In particular, it seems to appear systematically sub-luminous in diffuse X-rays when compared to other well-studied massive clusters observed in the same energy band (0.5–7.0\,keV), reflecting the variety of feedback mechanisms and physical conditions that govern the interaction between stellar winds and the surrounding ISM. The robustness of these conclusions relies on the careful quantification of instrumental and astrophysical contaminants detailed in Appendix~\ref{ap:xray_data} and on the observation-by-observation photon statistics compiled in Appendix~\ref{ap:table}.

\subsection{Absorption column densities and comparison with previous studies}
\label{sec:nh_discussion}

The CXE+NEI model adopted for region~\#1 yields $N_\mathrm{H} = 1.84 \times 10^{22}$~cm$^{-2}$ (frozen at the value preferred by the purely thermal fits), in agreement with values of $\sim 1.7$--$1.8 \times 10^{22}$~cm$^{-2}$ obtained with purely thermal models for the same region (Table~\ref{tab:spec}), and lower than the values of $\sim 2.0$--$2.2 \times 10^{22}$~cm$^{-2}$ reported by previous studies towards the cluster core \citep{Muno2006, Kavanagh2011, Haubner2025}. This lower absorption is physically consistent with the geometry of region~\#1: it lies to the south and outside the dense cluster core, in a direction where both the infrared-derived extinction maps and the HR map (Section~\ref{sec_hr}) indicate lower column densities. Purely thermal models are forced to absorb some of the CXE line flux into an effectively higher $N_\mathrm{H}$; once the CXE component is explicitly modelled, the remaining thermal continuum is better described by a genuinely lower column, making the $N_\mathrm{H} \simeq 1.84 \times 10^{22}$~cm$^{-2}$ physically meaningful rather than an artefact of model choice.

For region~\#2, the AP+PS model yields $N_\mathrm{H} = 2.24 \pm 1.20 \times 10^{22}$~cm$^{-2}$, consistent with the values of $\sim 2.0$--$2.2 \times 10^{22}$~cm$^{-2}$ reported towards the cluster core in previous studies \citep{Muno2006, Kavanagh2011, Haubner2025}. The somewhat higher $N_\mathrm{H}$ returned by AP+AP ($1.69 \times 10^{22}$~cm$^{-2}$) and the AP+NEI ($1.66 \times 10^{22}$~cm$^{-2}$) is a direct consequence of the harder intrinsic continuum produced by the PSHOCK component, which requires a correspondingly higher absorbing column to simultaneously reproduce the observed soft-band flux. The AP+PS value is therefore physically reliable and consistent with the independent extinction estimate towards the Wd\,1 core.

Otherwise, to understand the influence of warm/cold gas on the region, we used our infrared JWST/MIRI observations of Wd\,1 from the EWOCS project \citep{Guarcello2024}. These observations provide an opportunity to test whether a low-density cavity in the ISM, indicated by a deficit in infrared emission, independently confirms such an evacuated structure. 
\begin{figure*}[h!]
\includegraphics[width=18.3cm,angle=0]{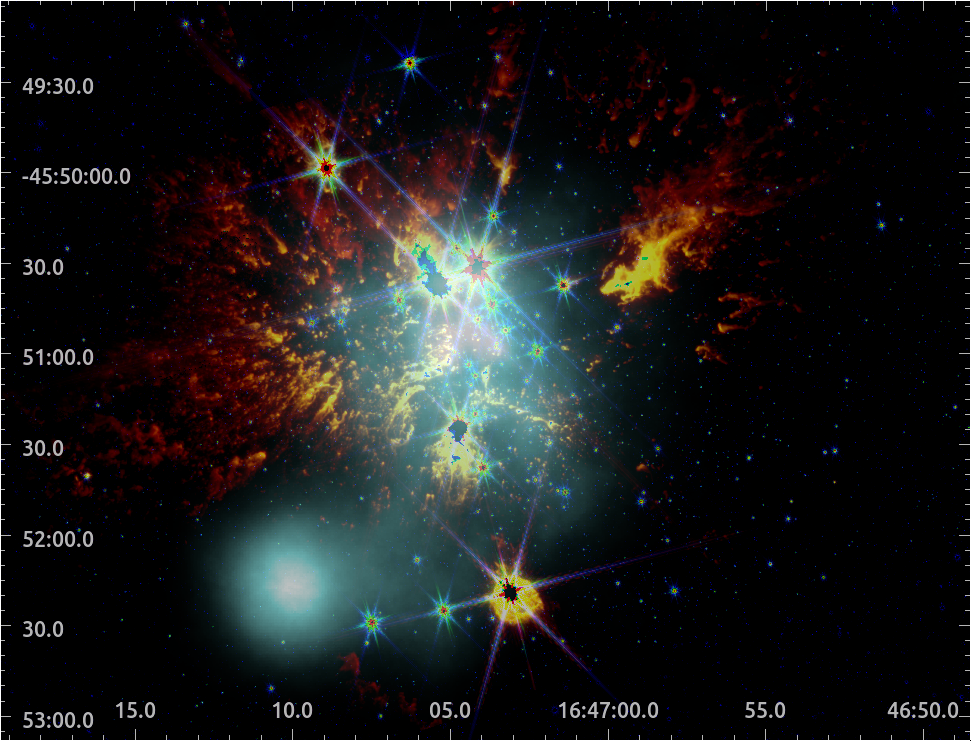}
\caption{Combined NIRCam/MIRI RGB image of Westerlund 1 (red: F1130W, green: F730W, blue: F444W). Diffuse X-ray emission in the 0.5-7.0 keV band is shown in smoked cyan colour.}
\label{img_xdif+miri}
\end{figure*}
In Figure \ref{img_xdif+miri}, we compared the spatial distribution of this mid-infrared emission with the diffuse X-ray emission in the 0.5–7.0 keV band. We found a weak negative spatial correlation of $r = -0.1891$ ($p \approx 1.69 \times 10^{-67}$) in the central region of Wd 1, calculated using the Pearson correlation coefficient \citep{Pearson1895} and its significance via the \texttt{pearsonr} function from \texttt{scipy.stats} \citep{Virtanen2020}. This result shows an anti-correlation between cold/warm dust and X-ray hot plasma, consistent with a scenario in which strong stellar winds from massive stars evacuate the central ISM. The infrared cold gas and warm dust structures are located outside the core, whereas the interior is dominated by diffuse X-ray-emitting plasma.

This process often creates large, low-density cavities that inhibit the formation of strong shocks with the surrounding medium. The efficiency of diffuse X-ray production depends critically on the shock regime that governs the thermalisation of stellar wind energy according to the expression: $T_{\rm sh} \simeq \frac{3}{16}\,\frac{\mu m_p}{k_B}\,v_w^2$, where $v_w$ is the terminal wind velocity, $\mu$ the mean molecular weight, $m_p$ the proton mass, and $k_B$ Boltzmann’s constant.
The corresponding diffuse X-ray luminosity can be approximated as $L_X \simeq n^2\,\Lambda(T)\,V$,
where $n$ is the characteristic gas density, $\Lambda(T)$ the optically thin cooling function, and $V$ the emitting volume. By normalising to the wind mechanical luminosity rate ($\dot{E}_{\rm w}$), one obtains a measure of the efficiency of converting stellar wind mechanical power into diffuse X-ray emission via
\begin{equation}
\eta_X \equiv \frac{L_X}{\dot{E}_{\rm w}} \propto \frac{n^2\,\Lambda(T)\,V}{\dot{E}_{\rm w}}.
\end{equation}

In radiative shocks ($\eta_X \lesssim 1$; \citealt{Stevens1992}), the post-shock gas cools efficiently, and a significant fraction of the mechanical power is converted into X-rays. Conversely, in adiabatic shocks ($\eta_X \ll 1$), typical of low-density environments, the energy is largely retained as thermal or kinetic energy, resulting in weak X-ray emission \citep{Lancaster2021b}. 

Processes that reduce the mean gas density, such as the radiation pressure or collective winds that evacuate the cluster core, directly suppress $L_X$ and therefore $\eta_X$, given the $L_X \!\propto\! n^2$ dependence. The relative influence of radiation and ram pressure can be quantified via
\begin{equation}
\Gamma \equiv \frac{P_{\rm rad}}{P_{\rm ram}} \simeq \frac{L_*}{c\,\dot{M}\,v_w},
\end{equation}
which shows that when $\Gamma \!\gtrsim\! 1$, radiation pressure dominates and drives gas outward, further lowering the density of the shocked plasma. Consequently, in compact, massive clusters with intense stellar feedback, such as Wd\,1, the conversion efficiency of mechanical wind energy into diffuse X-ray emission is intrinsically low, as most of the energy is expended in large-scale expansion and bulk motions rather than local radiative losses.\\ 

\begin{table}[ht]
\centering
\caption{Diffuse X-ray and stellar wind luminosities of Galactic massive star-forming regions.}
\label{tab:Lx_Lwind_refs_scaled}
\begin{tabular}{lccc}
\hline
Region & $L_X^{\rm diffuse}$  & $L_{\rm wind}$  & $L_X/L_{\rm wind}$  \\
        & [erg s$^{-1}$] & [erg s$^{-1}$] & [$10^{-4}$] \\
\hline
Westerlund 1        & $9.5\times10^{33}$ & $2.2\times10^{39}$ & $0.0043$ \\
Carina Nebula       & $2.0\times10^{35}$ & $5.0\times10^{38}$ & $4.00$ \\
NGC 3603            & $2.0\times10^{34}$ & $3.0\times10^{38}$ & $0.67$ \\
Cygnus OB2          & $4.2\times10^{34}$ & $1.5\times10^{38}$ & $2.80$ \\
Quintuplet Cluster  & $1.8\times10^{34}$ & $1.0\times10^{38}$ & $1.80$ \\
W49A                & $3.0\times10^{33}$ & $5.0\times10^{37}$ & $0.60$ \\
NGC 2024            & $2.0\times10^{31}$ & $5.0\times10^{35}$ & $0.40$ \\
RCW 38              & $7.0\times10^{32}$ & $1.0\times10^{37}$ & $0.70$ \\
W3 Main             & $5.0\times10^{32}$ & $1.0\times10^{37}$ & $0.50$ \\
Orion Nebula        & $5.5\times10^{31}$ & $1.0\times10^{36}$ & $0.55$ \\
\hline
\end{tabular}
\tablefoot{Diffuse X-ray luminosities ($L_X^{\rm diffuse}$), stellar wind luminosities ($L_{\rm wind}$), 
and their ratios ($L_X/L_{\rm wind}$, in units of $10^{-4}$)
for galactic massive star-forming regions. The list is sorted by column 3, from higher to lower L$_{\rm wind}$. 
Comparative regions are:
Westerlund 1: This work;
Carina Nebula: \cite{Townsley2011b,Sasaki2024};
NGC 3603: \cite{Moffat2002,Melena2008};
Cygnus OB2: \cite{Albacete-Colombo2023a},
Quintuplet Cluster: \cite{Rockefeller2004,Wang2006};
W49A: \cite{Tsujimoto2006,Smith2009};
NGC 2024: \cite{Ezoe2006};
RCW 38: \cite{Wolk2002,Fukushima2023};
W3 Main: \cite{Feigelson2008,Persi2008};
Orion Nebula: \cite{Guedel2008,Arthur2012}.}
\end{table}

\subsection{Efficiency of wind energy conversion into diffuse X-rays}
\label{sec:eta_discussion}

We used our X-ray observation of Wd\,1 as an empirical criterion to assess the efficiency with which mechanical wind energy is converted into diffuse X-ray emission. Two key quantities are compared in this work: (i) the total mechanical wind power, $L_{\rm w}$, estimated as $2.2\,(\pm0.6)\times10^{39}$~erg~s$^{-1}$, based on the observed WR population and wind evolution models specifically calibrated for Wd\,1 at an age of 4–5~Myr \citep{Harer2023}; and (ii) the total diffuse X-ray luminosity of the hot plasma is $9.5\,(\pm0.2)\times10^{33}$~erg~s$^{-1}$. The resulting kinetic-to-diffuse conversion efficiency, defined as $\eta = L_{\rm X}/L_{\rm w}$, is therefore $\eta \simeq 4.3\times10^{-6}$ for Wd\,1. This value is approximately two orders of magnitude lower than that reported for the Cygnus~OB2 association, $\eta = 2.8\times10^{-4}$, derived using an equivalent analysis methodology \citep{Albacete-Colombo2023a}. Table~\ref{tab:Lx_Lwind_refs_scaled} summarises the efficiency of stellar wind energy conversion into diffuse X-ray radiation across several massive star-forming regions. It should be noted, however, that differences in the treatment of point-source contamination and diffuse emission extraction introduce dispersion in the $L_{\rm X}$–$L_{\rm w}$ relationship. These factors may influence whether the observed trend remains linear or shows a turnover in the adiabatic regime, where the ICM has been partially evacuated from the core to the outer layers. The exceptionally low value of $\eta$ for Wd\,1 compared to other regions (Table~\ref{tab:Lx_Lwind_refs_scaled}) is physically plausible given its extreme compactness and the intense UV radiation field, which drives efficient ISM evacuation. The resulting low-density environment places Wd\,1 firmly in the adiabatic shock regime ($\eta_X \ll 1$ in the notation of Eq.~2), where most of the wind mechanical energy is retained as thermal or kinetic energy rather than being radiated away. Systematic differences in how the extraction regions are defined across studies may also contribute to the apparent spread; regions encompassing larger areas tend to include more contamination from background and unresolved sources, biasing $L_X$ upward. The robustness of our photon extraction against these effects is assessed in Appendices~\ref{sec:agn}, \ref{sec:foreground}, and \ref{sec:bkg}.

Finally, Wd\,1 stands out as a remarkable starburst cluster, not only because of its extremely large UV radiation and stellar winds from its population of massive stars, but also due to its low level of diffuse X-ray emission from shocked ISM. This provides valuable insights into how the ISM is restructured, thereby altering the physical conditions under which high-mass stars evolve and shaping the environments in which low-mass stars form.

\begin{acknowledgements}
We thank the anonymous referee for the careful reading of the manuscript, constructive comments, and suggestions that substantially improved the clarity and quality of this work. J.F.A.C. is a researcher of CONICET and professor at the National University of Rio Negro (UNRN) and acknowledges their support. This work was also supported by OCEANS Project No. 101183150, funded by the European Union.
The National Aeronautics and Space Administration also provided support for this work through Chandra Proposal 21200267 issued by the Chandra X-ray Center, which the Smithsonian Astrophysical Observatory operates for and on behalf of the National Aeronautics and Space Administration under contract NASA8-03060. CJKL gratefully acknowledges support from the International Max Planck Research School for Astronomy and Cosmic Physics at the University of Heidelberg in the form of an IMPRS PhD fellowship.
This study used these software packages: 
Numpy \citep{HarMilVan20}, matplotlib \citep{Hun07}, pySTARBURST99 \citep{Hawcroft2025}.
\end{acknowledgements}

\bibliographystyle{aa}
\bibliography{all_facu1_aa60124}

\begin{appendix}

\section{X-ray analysis}
\label{ap:xray_data}

Wd\,1 has a total stellar mass of 5$\times$10$^4$ M$_\odot$ and contains 10$^5$ young stars, most with masses between 0.3 and 3.0 M$_\odot$ \citep{Brandner2008, Andersen2017}. These low- and intermediate-mass stars are known to exhibit strong X-ray emission driven by magnetically active coronae \citep{Preibisch2005}. Here, we constructed an X-ray source density map based on the detected point sources. Although this map includes both stellar and instrumental components, it serves as a diagnostic tool to delineate the spatial boundaries of our analysis (see Fig.\,\ref{fig:img_fov}). Specifically, it confirms that the X-ray emission observed in the \textit{Chandra} dataset is spatially confined within an $8\times8$ arcminute field of view (FOV).

Evaluating instrumental limitations is essential to accurately detect and characterise the diffuse X-ray emission in Wd\,1, and to avoid biases and uncertainties when interpreting its properties.

\subsection{Detected point-like X-ray sources}
\label{sec:xsources}

The sensitivity and spatial resolution of the available observations intrinsically constrain the study of diffuse X-ray emission. In this work, we exploit the deepest \textit{Chandra}/ACIS-I dataset of the Wd\,1 region obtained to date, in which 5963 X-ray sources were detected within the $17\times17~\mathrm{arcmin}^2$ FOV \citep{Guarcello2024}. This catalogue represents the most complete census of X-ray sources currently available for Wd\,1, and provides a robust basis for deriving the radial distribution profile of the source population and assessing its relation to the extent of the diffuse emission.  

\begin{figure}[ht!]
\centering
\includegraphics[width=9.2cm,angle=0]{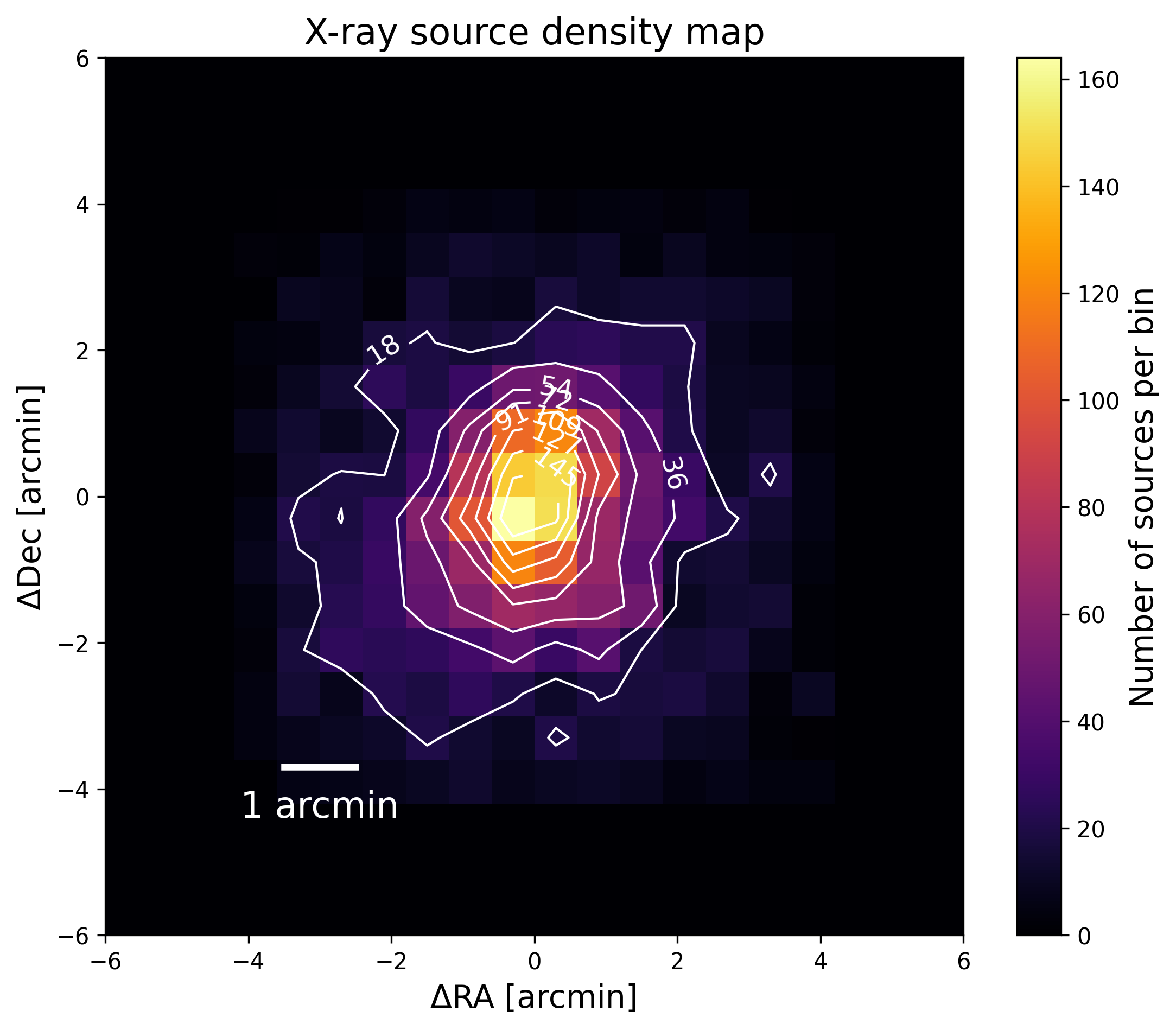}
\caption{\small 
X-ray source density map of the 4922 detected sources within an $8\times8$~arcmin$^2$ region centred at the stellar core of Wd\,1 (RA=16h:47m:04.53s, Dec=-45°:50':50.40").}
\label{ph_fov}
\end{figure}

For the analysis of the diffuse component, we selected a square region $8\times8~\mathrm{arcmin}^2$ centred on the cluster (see Figure~\ref{ph_fov}), encompassing more than twice the spatial extent of the detected X-ray sources. Within this field, 4922 sources were identified, where each source contributes different photon fluxes. This source distribution provides the foundation for constructing the radial density profile of the stellar population and evaluating its spatial correlation with the diffuse X-ray emission detected in the same region.  

We fit the radial distribution of the X-ray source density with a power-law model of the form $N(r) \propto r^{-\alpha}$, where $r$ is the projected distance from the cluster centre. The best fit is $\alpha \simeq 1.9 \pm 0.2$, which is consistent with a centrally concentrated population that decreases rapidly with radius. The power-law fit remains valid up to $r \sim 3$ arcminutes, beyond which the profile flattens and becomes indistinguishable from the background level. This result indicates that the majority of the X-ray sources in Wd\,1 are concentrated within $\sim3$ arcminutes ($\sim2.7$~pc at the assumed distance). 

\subsection{Smoothing the observation}
\label{sec:voronoi}

\begin{figure}[ht!]
\centering
\includegraphics[width=9cm,angle=0]{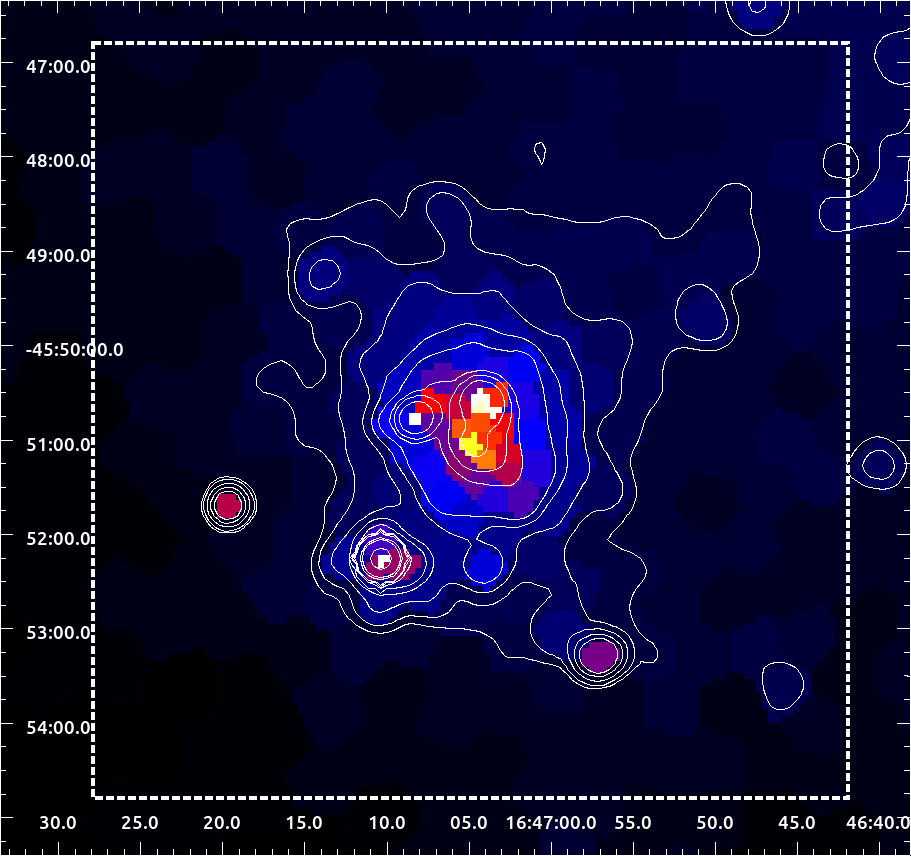}
\caption{\small 
Weighted Voronoi tessellation (WVT) map of the photon distribution with S/N = 49, constructed from the stacked 1~Ms dataset. The FOV is $8.0\times8.0$~arcmin$^2$, centred at RA=16:47:03.94, Dec=-45:51:10.09. The emission includes contributions from the resolved X-ray sources (4922), the unresolved source population, the diffuse X-ray component, and the instrumental background.}
\label{fig:smoothed_voronoi}
\end{figure}

From the stacked 1~Ms \textit{Chandra} observation, we constructed a 2D photon flux map in the $0.5$--$7.0$~keV energy range (see Fig.\,\ref{fig:smoothed_voronoi}). 
A total of 180,244 photons were detected in the 1\,Ms exposure time of the observation. However, based on the source detection analysis presented by \citet{Guarcello2023}, the total X-ray counts recorded in the selected $8 \times 8$~arcmin$^2$ field represent only $\sim28.5\%$ of this value. The remaining $\sim71.5\%$ of the observed signal cannot be attributed to the catalogued population but is due to a combination of instrumental background, galactic foreground stars, unresolved background star populations, extragalactic contamination, and, most importantly, diffuse X-ray emission.
This imbalance is a crucial aspect of the dataset: the observed X-ray signal is dominated by components other than the resolved stellar sources. It is important to note that the photon distribution is also affected by additional contributions, including $i:$ unresolved AGNs that remain undetected in the hard X-ray band (see Sect.\,\ref{sec:agn}); $ii:$ soft X-ray emission from foreground stars projected along the line of sight (see Sect.\,\ref{sec:foreground}); $iii:$ the residual instrumental background intrinsic to the detector (see Section\,\ref{sec:bkg}). All of these factors complicate the interpretation of the photon flux map and must be carefully considered to quantify their contribution to the truly diffuse X-ray component.

\subsubsection{The AGN background contribution}
\label{sec:agn}

The contamination from background AGNs on extended X-ray emission may introduce an excess of emission at X-ray energies. AGNs exhibit a highly heterogeneous spatial distribution, with surface densities ranging from $\sim$10--100~deg$^{-2}$ at $f_{2-10} \sim 10^{-14}$~erg~cm$^{-2}$~s$^{-1}$ to $\sim$1000~deg$^{-2}$ at $f_{2-10} \sim 10^{-16}$~erg~cm$^{-2}$~s$^{-1}$, following a steep $\log N$--$\log S$ relation \citep{Moretti2003, Lehmer2012}. Their clustering is characterised by a correlation length $r_0 \approx 6$--$10~h^{-1}$~Mpc \citep{Ebrero2009}, indicating moderate large-scale structure, while on arcmin-degree scales their distribution appears largely Poissonian with significant field-to-field variance. Importantly, AGNs contribute predominantly to the hard X-ray band due to their power-law spectra ($\Gamma \sim 1.5$--2.0; \citealt{Tozzi2006, Luo2017}), and are not expected to contribute significantly to the soft diffuse X-ray emission ($E \lesssim 2$~keV).
 
To estimate the contribution of the CXB, we can assume its emission to be statistically homogeneous across the entire FOV, as it is dominated by unresolved extragalactic sources. The extended background component was computed by extracting a 1~Ms spectrum from a sky region matching the area used for the diffuse emission analysis. Following \cite{Muno2006}, the average AGN contribution in the 2--7~keV band is modelled as a power law with a photon index $\Gamma = 1.5$ and an absorption column density of $N_{\rm H} = 2.2 \times 10^{22}$~cm$^{-2}$. This yields an observed count rate of $\sim3.0 \times 10^{-3}$~ph\,s$^{-1}$, corresponding to a 2.0--7.0~keV flux of $8.7 \times 10^{-14}$~erg\,cm$^{-2}$\,s$^{-1}$. This translates into an upper limit for the hard X-ray flux of $\sim\,13$\% of the total 2.0--7.0~keV diffuse X-ray flux and only $\sim1.4$\% in the broader 0.5--7.0~keV energy band (see Section~\ref{sec:spectral}).

\subsubsection{Foreground stellar contribution}
\label{sec:foreground}

The contamination from undetected foreground stars represents a more subtle challenge, since their spectral energy distributions closely resemble that of the diffuse X-ray emission under study. In particular, \citet{Kashyap2023} analysed the stacked spectrum of all individually detected foreground stars towards the Cygnus\,OB2 region and confirmed that their emission is consistent with a soft thermal plasma characterised by $kT = 0.77$~keV and $N_\mathrm{H} \approx 0.2 \times 10^{22}$~cm$^{-2}$. This implies that a non-negligible fraction of the diffuse X-ray signal observed in our study may be attributable to a foreground stellar population. 

To obtain an observational constraint for the case of Wd\,1, we followed the methodology of \citet{Getman2011}, who derived stellar surface densities for the Carina Nebula, which contains a total of  N=200,000 stars distributed over an area of $A = 1.42~\mathrm{deg^{2}} = 5,112~\mathrm{arcmin^{2}}$, yielding a surface density of $\rho_{2.3} = N/A \approx 39.1~\mathrm{stars~arcmin^{-2}}$ up to a distance of 2.3~kpc. Assuming a linear increase of contamination with distance, the scaling factor to the distance of Wd\,1 (4.2~kpc) is $f = 4.2/2.3 \approx 1.83$, giving a projected density of $\rho_{4.2} = \rho_{2.3} \times f \approx 71.4~\mathrm{stars~arcmin^{-2}}$. For the diffuse emission region in Wd\,1, which subtends a fixed projected area of $A_{Wd1} = 7.3~\mathrm{arcmin^{2}}$, the expected stellar contamination is $N_{cont} = \rho_{4.2} \times A_{Wd1} \approx 71.4 \times 7.3 \simeq 521$ stars. Furthermore, adopting a photon detection threshold of $2 \times 10^{-6}~\mathrm{ph~s^{-1}}$ for a 1~Ms \textit{Chandra}/ACIS exposure \citep{Guarcello2019cxo}, the expected contribution from each undetected source would be $N_{\gamma} = (2 \times 10^{-6}) \times 10^{6} = 2$ photons. Extrapolated to the approximately 521 potential undetected sources, this yields a total contribution of about 1,042 photons across the diffuse emission region. In comparison with the 12,300 photons detected in the diffuse X-ray spectrum of Wd\,1 (see Section \ref{sec:spectral}), this value represents only approximately 8.5\% of the total counts. We therefore infer that contamination by foreground stars constitutes only a minor contribution to the observed diffuse X-ray emission in Wd\,1 and does not exert a significant influence on the interpretation of its spectral characteristics.

\subsubsection{Undetected stellar contribution in Wd\,1}
\label{sec:bkg}

The star cluster Wd\,1 hosts about 10$^{5}$ stars distributed across its central region with a diameter of 3.5\arcmin \citep{Andersen2017}. Given its distance, a substantial fraction of low-mass members remain undetected, and their combined X-ray emission may contribute to the observed diffuse X-ray background. MARX simulations of a 1~Ms{} \textit{Chandra}/ACIS observation set a detection limit of $1.5 \times 10^{-16}$~erg\,cm$^{-2}$\,s$^{-1}$ \citep{Guarcello2019cxo}, illustrating the challenges in resolving the faintest cluster members and assessing their contribution to the diffuse emission. 
According to MARX simulations, detectable X-ray emission is expected from $\sim5030$ stars in the $0.8$–$2.0$\,M$_\odot$ mass range, of which only $\sim2100$ ($\sim42$\%) would be individually detected, leaving up to $\sim2900$ sources with $M < 1$\,M$_\odot$ below the sensitivity threshold \citep{Guarcello2019cxo}. Using the COUP X-ray luminosity function scaled to the distance and extinction of Wd\,1, we estimate a characteristic X-ray luminosity of $\sim1.3 \times 10^{29}$~erg\,s$^{-1}$ for these undetected stars, which represents an upper limit to their collective contribution. Considering the observed spatial distribution of X-ray sources in Wd\,1 (FWHM $\sim4$\arcmin), with 63.7\% and 87.7\% located within 1\arcmin\ and 3\arcmin\ of the cluster centre, respectively \citep{Guarcello2024}, we infer that up to $\sim2550$ undetected stars are likely situated within $\sim1.5$\arcmin\ of the diffuse X-ray emission region (see Section~\ref{sec:radprofile}). The integrated luminosity of this unresolved low-mass population is therefore constrained to be $\lesssim 3.3 \times 10^{32}$~erg\,s$^{-1}$, corresponding to only $\sim 3.1$\% of the total 0.5–7.0~keV diffuse X-ray luminosity (see Section~\ref{sec:spectral}). Thus, the contribution of faint undetected stars to the diffuse emission remains as a minor fraction of the overall diffuse X-ray luminosity in Wd\,1.

\section{Diffuse emission analysis}
\label{ap:analysis}

\subsection{Single observation analysis}
\label{sec:single}

We used the AE \textsc{ae\_construct\_regions} tool and the MARX software \citep{Davis2001} to generate individual PSF images from each observation’s aspect files. We analysed each observation independently, using the full source catalogue to extract point-source photons. To subtract point sources, intensity models are built and masks are drawn for all 4,922 detected sources using PSFs computed at 0.5, 1.0, 1.5, 2.7, and 4.0~keV. The \textsc{ae\_better\_masking} routine adaptively defines 99\% enclosed-energy masks, appropriate for faint sources where PSF wing losses are minimal. For bright sources, PSF wings near masked regions are inspected to avoid contamination from scattered photons. AE ensures that residual PSF-wing or scattered photons remain below the local background level \citep{Broos2012}.

The next step in point-source subtraction involves the AE task \textsc{ae\_make\_catalog}, which builds intensity models and generates source masks to exclude point-source contributions from the event file. This is achieved by simulating the X-ray beam at various off-axis angles and computing local PSFs across different energies. Given the crowded nature of Wd1, many sources have overlapping PSFs, requiring recalculation using the \textsc{ae\_better\_background} task. This routine adaptively refines aperture thresholds to enclose 99\% of the source energy, which is especially effective for faint, well-isolated sources. It also mitigates contamination from PSF wings, reducing the risk of source photon loss. For partially blended sources, AE models and subtracts residual light contamination from neighbouring sources using advanced iterative background correction algorithms \citep{Broos2012}. The photon-flux contribution of the source wings is approximately
\begin{equation}
\hskip 1.5cm \textsc{WINGS$_{ph}$ $\leqslant$ SRC\_CNTS $\times$ (1 - PSF$_{frac}$)}.
\end{equation}

To reliably assess the influence of source wings on the diffuse emission, it is essential to generate a realistic model of the instrumental background using data from the same observation. 

\subsection{Instrumental background contribution}
\label{sec:instrumental}

\begin{figure}[ht!]
\centering
\includegraphics[width=8.9cm,angle=0]{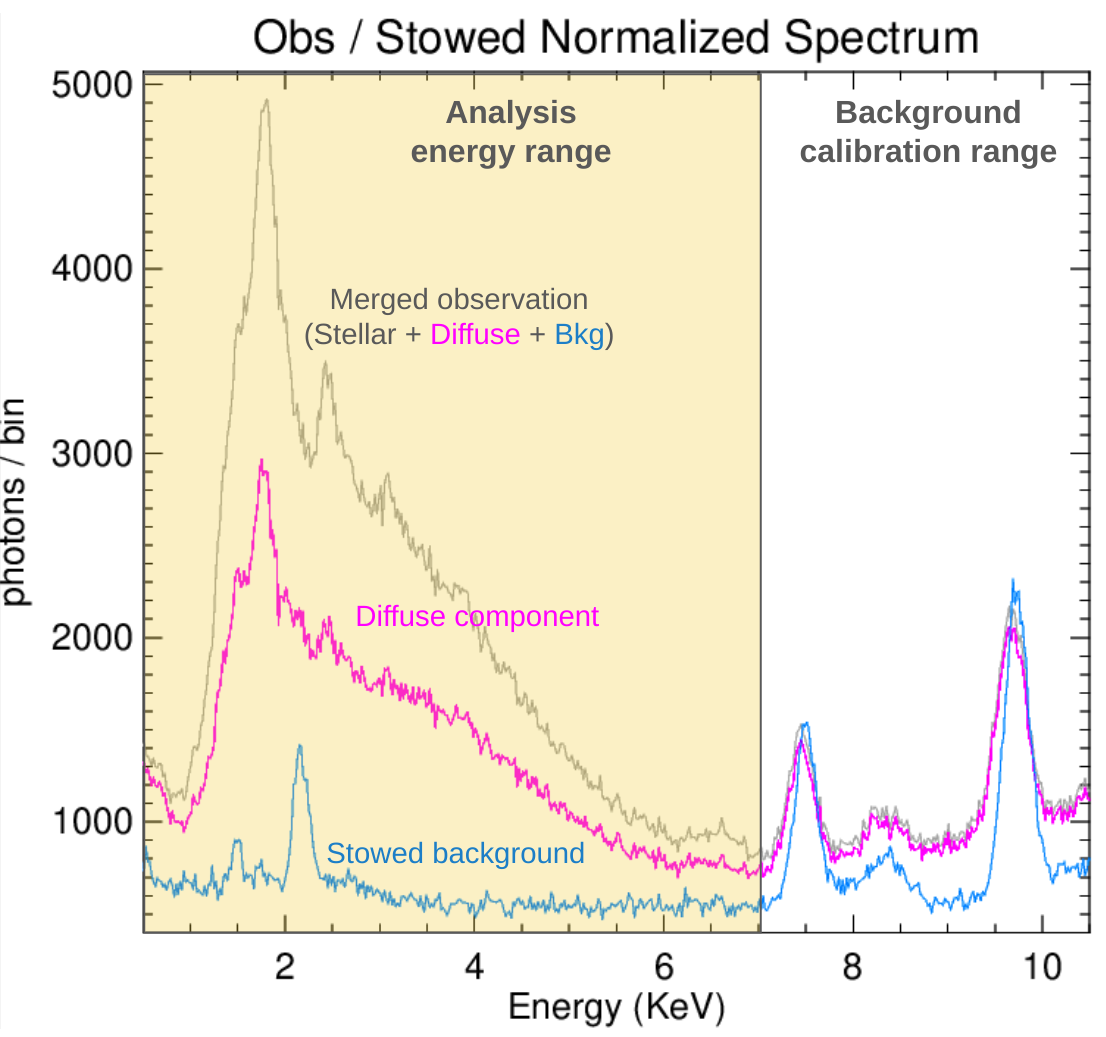}
\caption{Pulse-height distribution of events for the stacked observation (grey), the source-removed stacked file (magenta), and the normalised stowed background (cyan), matched in exposure using the instrumental emission lines at 7.47 and 9.8 keV. The sky shows no emission lines. The 2.15 keV line is also instrumental but was intentionally not used for the normalisation to test for its possible presence in diffuse X-ray spectra.}
\label{fig:OBs-Stw}
\end{figure}

Estimating the background for diffuse X-ray emission is challenging, as the signal extends across the entire detector and cannot be directly measured—even when it falls below the sensitivity threshold—due to contamination from non-local sources such as energetic particles from the solar system and radiation belts, as well as unresolved background sources such as AGNs. To mitigate this, we employed a standard ACIS calibration file to model the instrumental and particle background for each observation. Given that the ACIS background varies over time, it was necessary to adjust the background images individually, rather than relying on relative exposure times. Each background was scaled to match the spectral shape of its corresponding observation in the 7–12 keV band, where stellar emission is negligible \citep{Hickox2006}. In Figure \ref{fig:OBs-Stw}, we show the stacked photon spectrum used to calculate the depth of the stowed file. This assisted in creating new stowed exposure maps for the upcoming analysis stage. These maps were applied to each source position, in conjunction with mask-stowed event files. The resulting total background photon counts are listed in column 6 of Table\,\ref{tab:tab_obs}.

We determine the total background photons (DIFF$_{ph}$) at each source position by adjusting the background photons calculated by AE (BKG$_{ph}$) to give the number of photons matching the instrumental calibration data (STW$_{ph}$). This provides an accurate count of genuine instrumental photons for each detected source inside its PSF area:

\begin{equation}
\hskip 2.9cm \textsc{DIFF$_{ph}$ = BKG$_{ph}$ - STW$_{ph}$}.
\end{equation}

\begin{figure}[h!]
\centering
\includegraphics[width=9cm,angle=0]{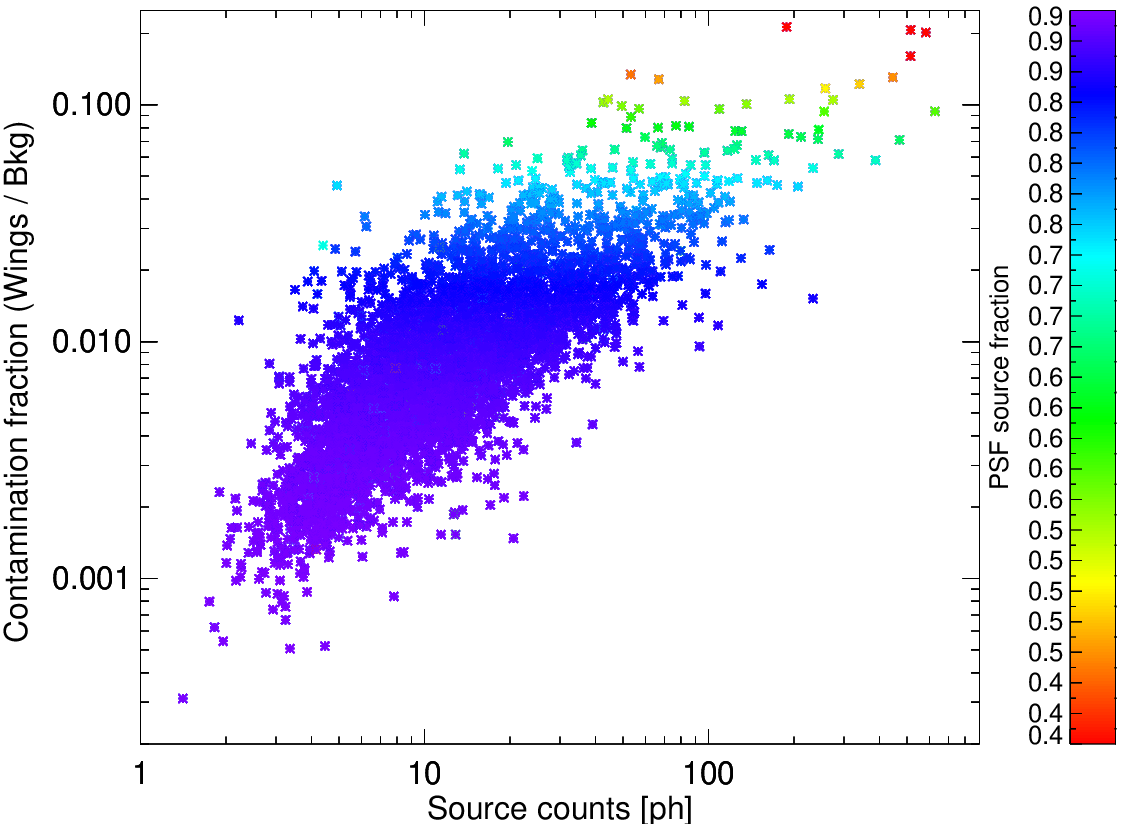}
\caption{\small X-ray source wings to the local background contamination (Wings$_{ph}$/Diff$_{ph}$) fraction as a function of the individual source counts (Src-counts). Sources with PSF fractions lower than 0.7 were redefined to a 1.2 arcsec source radius (see text).} 
\label{fig:img_w-b_cont}
\end{figure}

Accounting for our source detection procedure \citep{Guarcello2024} and the analysis steps \citep{Albacete-Colombo2023a} using the AE software, we characterised the photometric properties of each source to estimate the number of photons escaping their local PSFs. This enables a direct numerical assessment of how source wings impact the observed diffuse X-ray emission. Typically, the wings contribute less than 1\% to the local background, and only in rare cases, especially for bright sources (e.g. massive stars) can they reach up to 10\%. In these cases, we carefully inspect the masked regions to identify residual scattered photons. When necessary, we expand the standard 99\% PSF mask by a factor of up to 1.5 \citep{Townsley2011a}, ensuring that any remaining photons

We define the contamination fraction (FRAC$_{cont}$) in Equation A.3 as the ratio of photons from the source wings to the local diffuse emission at the same position. Figure\,\ref{fig:img_w-b_cont} shows that, among the 4,922 X-ray sources analysed, only 27 had PSF fractions between 0.4 and 0.5, with a maximum contamination fraction below 0.14 (14\%). Table\,\ref{tab:tab_obs} presents the contamination fractions for each observation, calculated as:
\begin{equation}
\hskip 1.5cm \textsc{FRAC$_{cont}$ $\leqslant$ (WINGS$_{ph}$ / DIFF$_{ph}$)} \hskip 1cm [in \,\%].
\end{equation}

Table\,\ref{tab:tab_obs} summarises the key statistics of sources, background, and diffuse emission for each observation. The most relevant result is that the median contamination fraction from source wings relative to the diffuse emission is 10.4\% ($\pm$0.7). This indicates that any subsequent analysis will remain unbiased, with uncertainties dominated by statistical errors rather than instrumental effects, ensuring a robust physical interpretation.

\subsection{Stacked analysis}

To improve the photon statistics, we combined all available source-free observations into a single stacked event file using the CIAO routine \textsc{merge\_all} \citep{Fruscione2014}, resulting in a total exposure of 1\,Ms. However, differences in satellite orientations introduced significant offsets in the stacked file: 0.0415 for \textsc{ra\_nom}, 0.0200 for \textsc{dec\_nom}, and 339.0 for \textsc{roll\_nom}. These values exceed the recommended tolerance thresholds for spectral analysis (0.0003 and 1.0, respectively). Consequently, while this misalignment prevents the reliable construction of response matrices for spectral analysis (see Section~\ref{sec:spectral}), it does not affect the application of smoothing techniques (see Section~\ref{sec:smooth}), which can be safely performed on both merged exposures and instrumental background files processed under the same configuration. See appendix \ref{ap:table} for further numerical details.

\section{Scattered X-rays from magnetar CXO\,164710.2-455217}
\label{sec:cxo}

\begin{figure}
    \centering
    \includegraphics[width=0.99\linewidth]{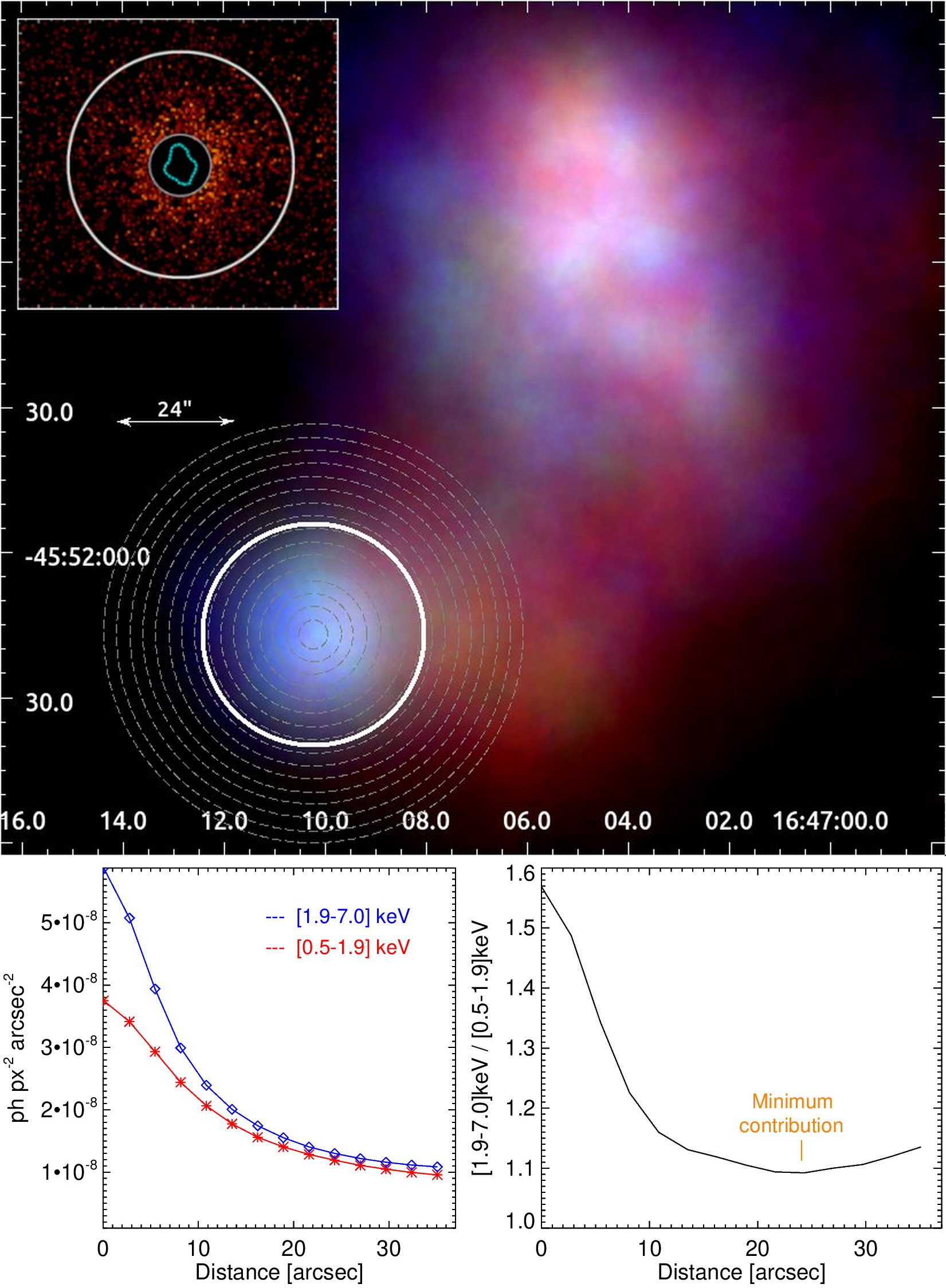}
    \caption{
Top: \textit{Chandra} colour image of the magnetar CXO~164710.2$-$455217 region (soft: red, medium: green, hard: blue). The dashed white circles show the concentric annuli used to compute the radial emission profiles, and the thick white circle marks the adopted exclusion boundary of 26.5~arcsec. The scale bar indicates 24~arcsec. The inset shows a close-up of the magnetar core, with the 95\% PSF polygonal region of the point source (cyan), the inner circle of 10~arcsec radius used to extract the core spectrum, and the outer circle of 26.5~arcsec delimiting the scattered-halo annulus.
Bottom left: Radial surface brightness profiles in the hard (1.9--7.0~keV; blue diamonds) and soft (0.5--1.9~keV; red asterisks) bands as a function of distance from the magnetar. Both profiles converge beyond $\sim$20~arcsec, indicating that the hard scattered contribution approaches the level of the soft diffuse background.
Bottom right: Ratio of the hard-to-soft surface brightness as a function of distance. The ratio reaches a well-defined minimum at $\sim$24~arcsec (0.49~pc at $d=4.2$~kpc; orange marker), where the magnetar-scattered hard emission is minimised relative to the soft diffuse component. We therefore adopt a conservative exclusion radius of 26.5~arcsec.}
    \label{fig:mag_radprof}
\end{figure}

This intense source is an anomalous X-ray pulsar classified as a magnetar associated with Wd\,1, based on its apparent proximity to the cluster and the low probability of a coincidental association \citep{Muno2006}. \textit{Swift} observations detected an X-ray burst from this source \citep{Campana2006} and \textit{XMM-Newton} observations were conducted both before and after the outburst \citep{Muno2007, Israel2007}. The interaction of its intense hard emission with interstellar dust produces a scattering halo \citep{Catura1983} whose spatial extent must be quantified before extracting the surrounding diffuse emission.

The magnetar emits predominantly hard X-rays ($>$2\,keV). Because scattering efficiency scales as $E^{-2}$, the scattered halo is intrinsically hard and concentrated within a few tens of arcseconds of the source. To determine the appropriate exclusion radius, we computed radial hard (1.9--7.0~keV) and soft (0.5--1.9~keV) surface brightness profiles using concentric annuli (Fig.~\ref{fig:mag_radprof}). The hard-to-soft ratio reaches a well-defined minimum at $\sim$24~arcsec (0.49~pc at $d = 4.2$~kpc), beyond which it flattens to a value representative of the ambient diffuse emission; we therefore adopted a conservative exclusion radius of 26.5~arcsec for the diffuse emission analysis.

\begin{figure}
    \centering
    \includegraphics[width=9.3cm, angle=0]{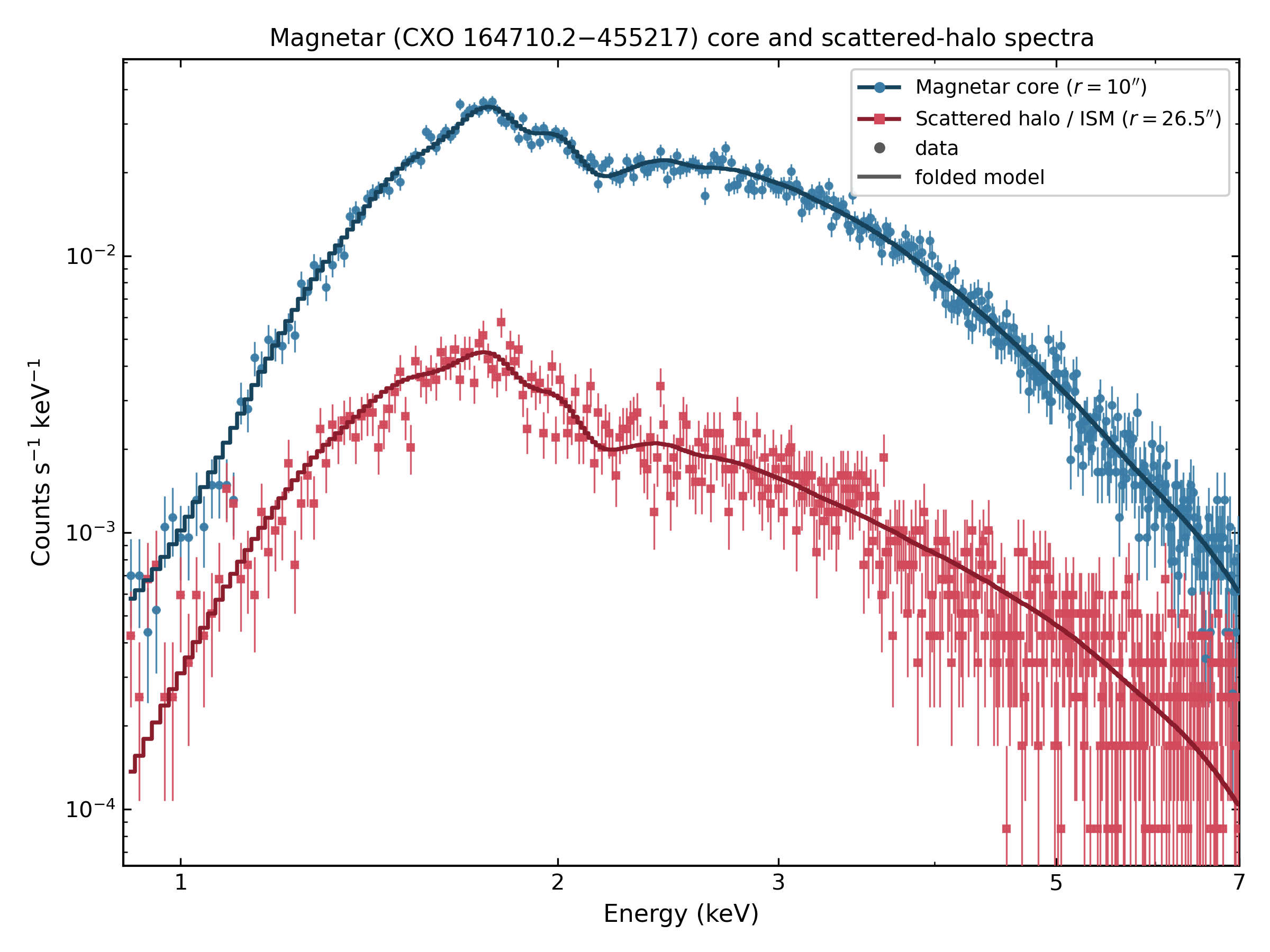}
    \caption{X-ray spectrum of the magnetar. The solid blue and red lines represent the spectrum and best-fitting model of the core and the scattered emission from the magnetar, respectively.}
    \label{fig:mag_spectra}
\end{figure}

We extracted the core spectrum from a circular region of 6.2~arcsec radius (1.5$\times$ the 95\% PSF extent) and the scattered halo spectrum from an annulus of inner/outer radii 10.0/26.5~arcsec, both in the 0.5--7.0~keV band, stacking all 36 observations. Dust scattering can underestimate the true absorption column by up to 25\% when unaccounted for \citep{Corrales2016}, so we fitted the two components independently (Fig.~\ref{fig:mag_spectra}) using an absorbed blackbody plus power-law model \citep{Mereghetti2008}.

For the core, the best-fit parameters are $N_\mathrm{H} = 1.86(\pm0.07)\times10^{22}$~cm$^{-2}$, $kT = 0.52\pm0.01$~keV, $\Gamma = 1.5\pm0.3$, with normalisations of $2.2(\pm0.1)\times10^{-5}$ and $9.1(\pm3.7)\times10^{-5}$~photons~cm$^{-5}$ for the blackbody and power-law respectively ($\chi^2/\mathrm{dof} = 460/417$, $\chi^2_\nu = 1.1$). The absorbed and unabsorbed 0.5--7.0~keV fluxes are $1.3(\pm0.2)\times10^{-12}$ and $2.4(\pm0.2)\times10^{-12}$~erg~cm$^{-2}$~s$^{-1}$, giving an absorption-corrected luminosity of $5.2(\pm0.2)\times10^{33}$~erg~s$^{-1}$ at the distance of Wd\,1 \citep{Muno2007}.

For the scattered halo, the same model yields $N_\mathrm{H} = 1.18(\pm0.08)\times10^{22}$~cm$^{-2}$, $kT = 0.47\pm0.01$~keV, $\Gamma = 1.3\pm0.3$, with normalisations of $1.8(\pm0.1)\times10^{-6}$ and $8.1(\pm0.5)\times10^{-5}$~photons~cm$^{-5}$ ($\chi^2/\mathrm{dof} = 506/417$, $\chi^2_\nu = 1.2$). The absorption-corrected luminosity in the 1.0--7.0~keV band is $0.45(\pm0.06)\times10^{33}$~erg~s$^{-1}$.

The combined core plus scattering luminosity is $L_X^{\rm core+scattering} = 0.57\times10^{34}$~erg~s$^{-1}$, which provides an upper limit to the magnetar's contamination of the diffuse X-ray emission reported for Wd\,1 in previous works.

\section{Quantitative results of extraction analysis}
\label{ap:table}

Here, in the table\,\ref{tab:tab_obs}, we give quantitative results and residuals of the AE analysis. From this table, we can describe the impacts of instrumental residual on our analysis, as it was presented in appendix \ref{ap:analysis}

\begin{table*}
\centering
\caption{Photon statistics, background, and diffuse components per observation.}
\label{tab:tab_obs}
\begin{tabular}{lcccccccc}
\hline
\multicolumn{1}{c}{Col 1} &
\multicolumn{1}{c}{Col 2} &
\multicolumn{1}{c}{Col 3} &
\multicolumn{1}{c}{Col 4} &
\multicolumn{1}{c}{Col 5} &
\multicolumn{1}{c}{Col 6} &
\multicolumn{1}{c}{Col 7} &
\multicolumn{1}{c}{Col 8} &
\multicolumn{1}{c}{Col 9}\\
\multicolumn{1}{c}{Obs.} &
\multicolumn{1}{c}{Sources} &
\multicolumn{1}{c}{Observation} &
\multicolumn{1}{c}{Source} &
\multicolumn{1}{c}{PSF wings} &
\multicolumn{1}{c}{Stowed} &
\multicolumn{1}{c}{Diffuse obs.} &
\multicolumn{1}{c}{Diffuse corr.} &
\multicolumn{1}{c}{Contamination}\\
\multicolumn{1}{c}{Id.} &
\multicolumn{1}{c}{in FOV} &
\multicolumn{1}{c}{(photons)} &
\multicolumn{1}{c}{(photons)} &
\multicolumn{1}{c}{(photons)} &
\multicolumn{1}{c}{(photons)} &
\multicolumn{1}{c}{(photons)} &
\multicolumn{1}{c}{(photons)} &
\multicolumn{1}{c}{fraction (\%)}\\
\hline
22316 &  2,166 & 27,517 &  7,722 &  902 & 9,771 & 18,098 &8,326 &10.8\\
22317 &  1,662 & 17,384 &  4,881 &  583 & 6,370 & 11,395 &5,025 &11.6\\
22318 &  1,717 & 18,265 &  5,055 &  588 & 6,332 & 12,050 &5,718 &10.3\\
22319 &  2,333 & 32,031 &  9,188 & 1,068 &11,426 & 20,877 &9,450 &11.3\\
22320 &  2,143 & 26,793 &  7,540 &  853 & 9,409 & 17,449 &8,040 &10.6\\
22321 &  2,052 & 25,747 &  6,980 &  738 & 9,281 & 17,215 &7,933 & 9.3\\
22977 &  2,099 & 25,850 &  7,249 &  840 & 9,277 & 17,042 &7,764 &10.8\\
22978 &  1,607 & 15,922 &  4,389 &  486 & 6,186 & 10,685 &4,498 &10.8\\
22979 &  1,493 & 14,586 &  3,949 &  444 & 5,519 &  9,805 &4,285 &10.4\\
22980 &  1,583 & 15,875 &  4,493 &  499 & 5,988 & 10,472 &4,483 &11.1\\
22981 &  1,506 & 14,501 &  3,832 &  431 & 5,393 &  9,973 &4,579 & 9.4\\
22982 &  1,306 & 12,138 &  3,465 &  383 & 4,297 &  7,890 &3,592 &10.7\\
22983 &  1,730 & 18,177 &  4,883 &  555 & 6,691 & 12,277 &5,585 & 9.9\\
22984 &  1,495 & 13,811 &  3,845 &  422 & 5,129 &  9,272 &4,142 &10.2\\
22985 &  1,680 & 18,567 &  4,856 &  550 & 6,775 & 12,579 &5,803 & 9.5\\
22986 &  1,310 & 12,602 &  3,516 &  389 & 4,448 &  8,338 &3,889 &10.0\\
22987 &  1,561 & 15,774 &  4,280 &  461 & 5,960 & 10,674 &4,713 & 9.8\\
22988 &  1,290 & 11,643 &  3,189 &  374 & 4,327 &  7,825 &3,497 &10.7\\
22989 &  1,461 & 14,574 &  3,936 &  443 & 5,623 &  9,922 &4,298 &10.3\\
22990 &  1,618 & 16,936 &  4,716 &  545 & 5,951 & 11,113 &5,161 &10.6\\
23272 &    952 &  8,145 &  2,203 &  236 & 2,990 &  5,490 &2,499 & 9.5\\
23279 &  1,866 & 21,100 &  5,964 &  671 & 7,373 & 13,760 &6,386 &10.5\\
23281 &  1,880 & 21,678 &  6,138 &  719 & 7,642 & 14,135 &6,492 &11.1\\
23287 &  2,035 & 23,884 &  6,644 &  754 & 8,540 & 15,791 &7,250 &10.4\\
23288 &  1,846 & 20,493 &  5,570 &  632 & 7,246 & 13,687 &6,440 & 9.8\\
24827 &  1,573 & 16,033 &  4,469 &  538 & 6,139 & 10,564 &4,424 &12.2\\
24828 &  1,575 & 16,067 &  4,283 &  450 & 6,255 & 10,922 &4,666 & 9.7\\
25051 &  1,906 & 21,303 &  5,702 &  633 & 7,719 & 14,385 &6,665 & 9.5\\
25055 &  1,726 & 19,912 &  5,102 &  539 & 7,545 & 13,639 &6,093 & 8.9\\
25057 &  1,625 & 15,628 &  4,484 &  504 & 6,122 & 10,323 &4,200 &12.0\\
25058 &  1,685 & 17,999 &  4,947 &  556 & 6,221 & 11,967 &5,745 & 9.7\\
25073 &  1,952 & 22,685 &  6,050 &  691 & 8,565 & 15,403 &6,837 &10.1\\
25096 &  1,253 & 11,782 &  3,215 &  375 & 4,372 &  7,866 &3,493 &10.8\\
25097 &  1,524 & 15,453 &  4,096 &  481 & 5,908 & 10,502 &4,593 &10.5\\
25098 &  1,647 & 17,142 &  4,756 &  540 & 6,140 & 11,494 &5,353 &10.1\\
25683 &  1,623 & 16,852 &  4,657 &  554 & 5,884 & 11,162 &5,277 &10.5\\
\hline
\end{tabular}
\tablefoot{Col 1: is the observation identification number. Col 2: the removed sources in the event file of the observation. Col 3: the total of X-ray photons in the file event. Col 4: number of photons from the sources in the observation. Col 5:  number of remaining photons from the source wings. Col 6: number of instrumental background photons for the observation. Col 7 = Col 3 - Col 4, gives the total of diffuse (source removed) photons. Col 8 = Col 7 - Col 6, shows the source and stowed background-corrected photons. Col 9 (see equation A.3) corresponds to the ratio between the total wings and the diffuse X-ray photons for each observation.}
\end{table*}

\end{appendix}

\end{document}